\documentclass[pdflatex,sn-mathphys-ay]{sn-jnl}

\usepackage{numprint}
\newcommand{\np}[1]{\numprint{#1}}
\newcommand{\rev}[1]{{#1}}

\usepackage{graphicx}%
\usepackage{multirow}%
\usepackage{amsmath,amssymb,amsfonts}%
\usepackage{amsthm}%
\usepackage{mathrsfs}%
\usepackage[title]{appendix}%
\usepackage{xcolor}%
\usepackage{textcomp}%
\usepackage{manyfoot}%
\usepackage{booktabs}%
\usepackage{algorithm}%
\usepackage{algorithmicx}%
\usepackage{algpseudocode}%
\usepackage{listings}%
\usepackage{subcaption}
\usepackage{siunitx} 
\usepackage{rotating}
\usepackage{tcolorbox}
\tcbuselibrary{listings,breakable}
\usepackage{float}
\usepackage{placeins}

\usepackage{threeparttable}
\theoremstyle{thmstyleone}%
\theoremstyle{thmstyletwo}%

\theoremstyle{thmstylethree}%

\begin{document}


\title{Large Language Models, Encoder Architectures and Hybrid Approaches for Patent Classification}



\author[1,2]{\fnm{Lorenzo} \sur{Emer}}\email{lorenzo.emer@santannapisa.it}
\author[3]{\fnm{Marco} \sur{Lippi}}\email{marco.lippi@unifi.it}
\author[1,4]{\fnm{Andrea} 
\sur{Mina}}\email{andrea.mina@santannapisa.it}
\author*[1,5]{\fnm{Andrea} \sur{Vandin}}\email{andrea.vandin@santannapisa.it}


\affil*[1]{\orgdiv{Institute of Economics and L'EMbeDS}, 
           \orgname{Scuola Superiore Sant’Anna}, 
           \orgaddress{\street{Piazza Martiri della Libertà, 33}, 
                       \city{Pisa}, \postcode{56127}, 
                       \country{Italy}}}

\affil[2]{\orgdiv{Department of Computer Science}, 
          \orgname{University of Pisa}, 
          \orgaddress{\street{Largo B. Pontecorvo 3}, 
                      \city{Pisa}, \postcode{56126}, 
                      \country{Italy}}}

\affil[3]{\orgdiv{Department of Information Engineering}, 
          \orgname{University of Florence}, 
          \orgaddress{\street{Via di Santa Marta 3}, 
                      \city{Florence}, \postcode{50139}, 
                      \country{Italy}}}

\affil[4]{\orgdiv{Centre for Business Research}, 
          \orgname{University of Cambridge}, 
          \orgaddress{\street{11–12 Trumpington Street}, 
                      \city{Cambridge}, \postcode{CB2 1QA}, 
                      \country{United Kingdom}}}

\affil[5]{\orgdiv{DTU Compute}, 
          \orgname{Technical University of Denmark}, 
          \orgaddress{\street{Anker Engelunds Vej 101}, 
                      \city{Kongens Lyngby}, \postcode{2800}, 
                      \country{Denmark}}}


\abstract{
Automated patent classification is essential for organizing technological knowledge and constructing indicators of technological change, specialization, and leadership. 
To identify the relative strengths and weaknesses of popular state-of-the-art approaches to this problem, we perform a controlled comparison of patent-specific encoders and open-weight local LLMs for hierarchical multi-label Cooperative Patent Classification (CPC). We find that the best-performing encoder (task-adapted BERT-for-Patents) outperforms the best-performing LLM (fine-tuned Qwen3.5-9B), while requiring one to two orders of magnitude less energy for inference. The two model families show complementary capabilities, which we leverage through a hybrid pipeline that routes patents with the highest encoder uncertainty to the LLM, yielding significant gains on this subset. We also find that across models, classification errors are especially pronounced in CPC categories that are cross-cutting or semantically broad --such as Section Y-- and that they have substantial consequences for technology mapping and country and assignee rankings.
The analysis covers predictive and hierarchical performance, computational cost and energy consumption, external validation on EPO patents, and the propagation of classification errors into downstream technological indicators, with implications for automated patent classification procedures used by patent offices, technology analysts, and scientometric researchers.  
}




\maketitle
  
\section*{Introduction}\label{sec1}

Patent classification is a foundational task in the organisation and analysis of technological information. In scientometrics and patentometrics, automated classification is used to map technological domains, trace innovation trajectories, and study technological change and emergence \citep{Haghighian2022PatentNet,Kim2020PatentClusteringDeepEmbeddings,Lu2024XLMR,Oh2020PatentClassificationLinkPrediction}. Accurate assignment of schemes such as the IPC and CPC is also important for patent offices, legal practitioners, technology analysts, and policymakers, supporting prior-art search, patent landscaping, and the monitoring of emerging technologies \citep{Krestel2021Survey,Lin2025Review}. The growth of patent filings and the increasing granularity of classification systems have therefore intensified the need for scalable methods capable of processing large, heterogeneous, multi-label, and hierarchically structured patent collections.

Automated classification remains relevant even when expert-assigned codes are available. Existing classifications may be incomplete, delayed, or inconsistent across patent offices and time periods \citep{Lafond2019PatentClassificationDynamics}; scientometric applications may require large corpora to be reprocessed under common criteria \citep{Choi2021EmergingTechnologiesPatentML}; and automated methods can identify relevant but unassigned categories in emerging or cross-domain technologies \citep{Lobo2019InventiveNoveltyClassification}. The objective is therefore not to replace examiner-assigned classifications, but to support large-scale analytical workflows and improve the consistency and coverage of patent-based indicators.

From a technical point of view, automated patent classification has progressed from bag-of-words representations and classical machine learning \citep{Chen2012ThreePhase,Wu2010Hybrid}, through neural architectures \citep{Li2018DeepPatent,Haghighian2022PatentNet}, to transformer-based language models. Fine-tuned encoders such as BERT, SciBERT, BERT-for-Patents, and PatentSBERTa provide strong and computationally efficient supervised solutions \citep{Lee2020PatentBERT,Bekamiri2024PatentSBERTa}. Large language models (LLMs), by contrast, can perform classification through zero-shot or few-shot prompting and can incorporate external knowledge through retrieval-augmented generation \citep{NEURIPS2020_1457c0d6,Bommasani2021FoundationModels,NEURIPS2020_6b493230}. Their broad pre-training has motivated the expectation that they may be particularly effective for rare, weakly represented, or semantically complex categories. However, evidence from text classification also shows that smaller task-adapted encoders may outperform prompted LLMs when sufficient labelled data and supervised optimisation are available \citep{edwards-camacho-collados-2024-language}.

Recent studies provide mixed evidence on the effectiveness of LLMs for technical, large-scale, and multi-label patent classification \citep{Kamateri2024AI,Rafieian2025LLM}. Standard supervised classifiers, however, tend to favour frequent categories under strongly imbalanced label distributions. The two families also differ substantially in latency, memory requirements, and energy consumption \citep{niu2025energy}. More broadly, scientometric research has stressed that automated systems should be evaluated not only in terms of predictive accuracy, but also with respect to bias, robustness, and the validity of the indicators they produce \citep{ThelwallKurt2025ChatGPTBias,Thelwall2025AIResearchQualityOpinion,ThelwallYaghi2025PeerReviewPrediction,Schmitt2024PatentQualityLLM}.

Therefore, several important questions remain unresolved. First, there is no controlled comparison of strong encoder and LLM approaches under comparably developed training and inference conditions, making it difficult to establish which family provides the most reliable basis for large-scale patent classification. Second, their complementarity remains insufficiently understood: existing studies do not determine whether encoders and LLMs fail on different patents, or whether selective combination can improve difficult cases. Third, most classification studies stop at predictive metrics and do not examine how model-specific errors propagate into scientometric indicators, including estimates of technological activity and rankings of countries and assignees.

We address these gaps through a controlled comparison of patent-domain encoders and open-weight local LLMs for hierarchical multi-label CPC classification. The analysis covers standard and long-tail-aware encoder training, zero-shot and few-shot LLM prompting, retrieval augmentation, and LoRA fine-tuning. We then combine the strongest encoder and LLM through an uncertainty-based hybrid system that routes only patents for which the encoder is least confident. The comparison includes predictive and hierarchical performance, computational cost and energy consumption, external validation on EPO patents, and the propagation of classification errors into technology counts and actor rankings.

The study makes both a methodological and an empirical contribution to scientometrics and patent-based research by showing how the choice of the model affects not only classification performance, but also the reliability of downstream indicators.

From a methodological perspective, we provide a controlled comparison of strong encoder and LLM configurations, examine the effects of task-specific adaptation, and identify the conditions under which the two model families complement each other. The adapted BERT-for-Patents encoder outperforms the LoRA-fine-tuned
Qwen3.5-9B model on the main predictive metrics while requiring substantially less energy at inference. At the same time, the hybrid system identifies a
subset of highly uncertain patents for which the LLM provides substantial gains.

From an empirical and patentometric perspective, we identify a persistent failure regime in semantically broad and cross-cutting CPC categories, particularly Section~Y. These categories remain substantially harder to classify across encoder, LLM, and hybrid systems. Their difficulty is associated not only with label frequency, but also with greater semantic breadth and weaker within-class cohesion, even after
accounting for training frequency, test support, and CPC section. The resulting errors propagate into distortions in estimated technology counts and shifts in
country and assignee rankings.
We further show that predictive accuracy and downstream ranking reliability are not necessarily tightly coupled. Although the LoRA-adapted LLM achieves lower
Micro-F1 than the encoder, the country and assignee rankings derived from its predictions are numerically closer to the gold-standard rankings across all examined. This indicates that the aggregate predictive performance of certain models does not necessarily translate proportionally into stable downstream patentometric indicators.

These findings have direct implications for patent offices, technology analysts, and scientometric researchers. Aggregate measures such as Micro-F1 and Macro-F1 are insufficient when predicted classifications are subsequently used to construct indicators. Model selection should also consider domain-specific error patterns, the semantic structure of the classification system, computational requirements, and the reliability of the particular metrics and rankings the model is intended to support.

\section*{Related work}
\subsection*{From traditional pipelines to encoder-based models}

Early work on automatic patent classification relied on classical text classification pipelines in which patent documents were represented using bag-of-words (BoW) features and classified with supervised learners such as support vector machines (SVMs) \citep{Wu2010Hybrid} or $k$-nearest neighbours (KNNs) \citep{murata2005knn}.
Term Frequency–Inverse Document Frequency (TF--IDF) has been the dominant BoW scheme for encoding titles, abstracts, or claims into high-dimensional sparse vectors.
These representations were typically used for multi-label prediction of IPC or CPC codes using linear classifiers.
Representative approaches include multi-phase and hybrid pipelines combining TF--IDF, feature selection, and SVM- or KNN-based classifiers, achieving moderate performance on collections of tens of thousands of patents
\citep{Chen2012ThreePhase,Wu2010Hybrid}.
Despite their scalability and simplicity, BoW-based methods are fundamentally limited by their inability to capture semantic similarity beyond lexical overlap and by their sensitivity to vocabulary drift in rapidly evolving technological domains \citep{turney2010frequency}.

The introduction of distributed word representations marked a transition from sparse BoW features to dense, low-dimensional embeddings.
Word2vec- and fastText-style embeddings trained on large patent corpora enabled more semantically meaningful representations of technical terminology and were combined with neural architectures such as convolutional neural networks (CNNs), recurrent neural networks (RNNs), gated recurrent units (GRUs), and long short-term memory (LSTM) networks \citep{xiao2018patent, risch2019domain}.
CNN-based models, in particular, demonstrated improvements over BoW+SVM pipelines by capturing local $n$-gram patterns in patent text
\citep{Li2018DeepPatent}.
Subsequent work extended these architectures to multi-label and hierarchical classification settings
\citep{Haghighian2022PatentNet,Nemati2024Semantic}.
Comparative evaluations on real-world patent datasets confirm that deep learning models such as TextCNN and TextRCNN generally outperform classical methods, although their effectiveness depends strongly on dataset size and label distribution \citep{xu2024multilabel_evaluation}.
Nevertheless, these models relied on static word embeddings, limiting their ability to handle polysemy, contextual variation, and evolving terminology, and often struggled with the length and structural heterogeneity of patent documents \citep{yang2016hierarchical}.

The introduction of transformer-based language models marked a substantial advance in patent classification.
Fine-tuning architectures such as BERT and its variants on patent corpora led to significant performance gains across single-label and multi-label classification tasks and across multiple levels of the IPC and CPC hierarchies
\citep{Lee2020PatentBERT}.
Domain-adapted variants trained on large-scale patent collections further improved performance by aligning the pretraining distribution with patent-specific language and stylistic conventions
\citep{Bekamiri2024PatentSBERTa, srebrovic2020bertpatents}.
As a result, encoder-based classifiers have become a de facto standard for automated IPC and CPC assignment, owing to their strong performance on frequent labels and their computational efficiency at inference time \citep{Lu2024XLMR}.
More recent work has explored enhanced architectures that incorporate temporal, hierarchical, and multi-source information, such as memory-augmented frameworks that model evolving IPC semantics and historical patent trends, yielding further improvements over standard transformer baselines \citep{xu2025memory_patent}.

From a patentometrics perspective, encoder models are attractive because they scale efficiently to millions of documents and can be seamlessly integrated into large indicator pipelines \citep{Krestel2021Survey}.
However, prior studies also note their sensitivity to label imbalance and their tendency to optimize performance on dominant subclasses \citep{pujari2021multitask, nam2014large}.
This limitation is particularly evident in real-world settings, where long-tailed label distributions and noisy IPC assignments remain a persistent challenge even for advanced deep learning models, raising concerns about their suitability for analyzing emerging or weakly represented technological areas
\citep{Krestel2021Survey,Lin2025Review,Kamateri2024AI}.

\subsection*{Large language models and prompt-based classification}

More recently, LLMs have been explored as an alternative paradigm for patent classification.
Unlike encoder-based systems, LLMs can perform zero-shot and few-shot classification by leveraging instruction-following capabilities acquired during large-scale pretraining.
Initial studies show that LLMs can assign plausible labels without task-specific fine-tuning and that performance can be improved through prompt engineering, few-shot exemplars, and lightweight retrieval augmentation
\citep{Yoshikawa2024Summaries, Rafieian2025LLM, xiong2025scalable}.
In low-resource settings, LLM-based approaches leveraging in-context learning have been shown to outperform traditional machine learning and deep learning baselines while providing improved interpretability \citep{yang2026gptpls}.
Similarly, retrieval-augmented few-shot frameworks combined with contrastive representation learning have demonstrated strong performance gains, particularly for multi-label classification in interdisciplinary domains \citep{zheng2026contrastive_patent}.

Applications of LLMs to patent analysis are still relatively limited in number and scope. Parallel discussions in scientometrics emphasize a similar pattern: LLMs are increasingly explored as assistive tools for evaluation and mapping tasks, but their reliability, bias, and resource requirements remain open concerns \citep{Thelwall2025AIResearchQualityOpinion, Sun2025LLMPeerReview, Cantone2025DisciplinarySimilarityLLM, Schmitt2024PatentQualityLLM}.
Beyond patent analysis, a growing body of work investigates the broader role of LLMs in automating scholarly evaluation processes, highlighting both their transformative potential and unresolved challenges related to robustness, reasoning, and human oversight \citep{zhuang2025llm_review_survey, lin2023automated_review}.
While recent studies demonstrate promising performance in specific settings—often by integrating LLMs into hybrid or assisted workflows such as active learning—most contributions remain task- or domain-specific and stop short of providing systematic, large-scale benchmarking across datasets, prompting strategies, and evaluation dimensions \citep{Kamateri2024AI,xiong2025scalable}.
Moreover, despite their relevance for operational deployment in patent offices and research institutions, the computational cost and energy footprint of LLM-based approaches are rarely analyzed as evaluation criteria.

\subsection*{Challenges, Measurement Error, and Responsible Use of Automated Patent Classification}

Patent classification remains intrinsically challenging. Patent documents are long, heterogeneous, and densely technical; classification systems are highly granular and hierarchical; and expert examiners themselves exhibit non-negligible disagreement in code assignment \citep{burke2007measuring}. As a result, automated classifiers often achieve relatively strong aggregate performance  metrics, while remaining substantially weaker on rare and fine-grained subclasses \citep{Krestel2021Survey,Lin2025Review,Kamateri2024AI}. These limitations have motivated research on ensemble methods, explainability, LLM-based classification, and hybrid systems \citep{Kamateri2023Ensemble,Rafieian2025LLM,xiong2025scalable}.

More broadly, the literature has already identified several requirements for the responsible use of automated classification in scientometric applications: models should be evaluated across different operating settings, their scalability and computational and environmental costs should be considered, and their consequences should be assessed beyond conventional predictive metrics \citep{hicks2015leiden,waltman2016review,schwartz2020green}. \citet{weber2020supervised}, for example, show that evaluation should reflect the intended downstream use, because different error types have different consequences across scientometric applications. \citet{robinsongarcia2024errors} further demonstrate that classification choices can introduce substantial measurement variability in bibliometric indicators.

Important gaps, nevertheless, remain. Existing studies provide limited evidence on how strong encoders, LLMs, and hybrid systems behave under comparable patent-classification settings; how their accuracy gains relate to inference cost, scalability, and energy consumption; and how their model-specific errors propagate into downstream indicators of technology mapping and country/assignee rankings. This is precisely where our contribution lies. We compare the two model families not only in terms of predictive, hierarchical, and rare-label performance, but also with respect to computational efficiency and energy use. We then examine whether classification errors are systematically more severe in semantically broad or cross-cutting CPC categories and whether they propagate into distortions in technology counts and country and assignee rankings. In this way, we connect document-level classification performance to the reliability of the indicators derived from predicted labels.

\section*{Methodology}
\label{sec:methodology}

\begin{figure*}[t!]
\centering
\includegraphics[width=0.9\linewidth]{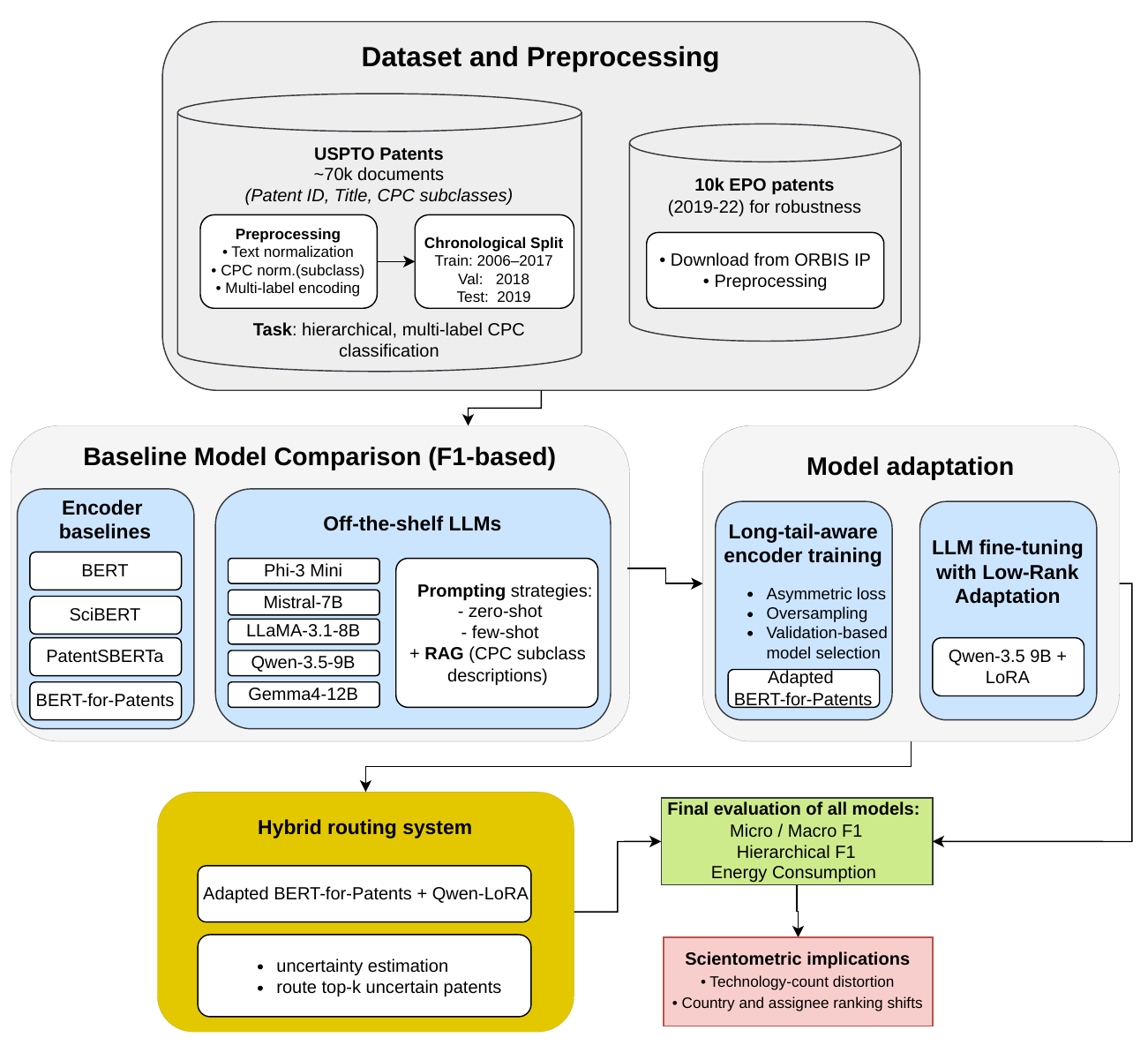}
\caption{Experimental pipeline for hierarchical multi-label CPC patent classification. Approximately \np{70000} USPTO patents are chronologically split into training (2006--2017), validation (2018), and test (2019) sets. The study first compares supervised encoder baselines (\rev{BERT, SciBERT, PatentSBERTa, and BERT-for-Patents}) against instruction-tuned LLMs under multiple prompting strategies, including zero-shot, few-shot, and retrieval-augmented prompting. The strongest encoder, \rev{BERT-for-Patents}, is subsequently enhanced using long-tail-aware training strategies, while the best-performing LLM, \rev{Qwen3.5-9B}, is adapted through \rev{a targeted LoRA hyperparameter sweep}. The resulting mitigated encoder and LoRA-adapted LLM are then combined through uncertainty-based hybrid routing, in which uncertain patents are selectively delegated to the LLM for refinement. Final evaluation includes flat, hierarchy-aware, rare-label, and energy-efficiency metrics, \rev{together with an analysis of classification-error propagation into technology counts and country and assignee rankings}. Findings are additionally validated on an external EPO patent dataset.}
\label{fig:methodology}
\end{figure*}

This study compares supervised encoder models and large language models (LLMs) for CPC subclass-level multi-label patent classification, with particular attention to predictive performance, long-tail behaviour, computational efficiency, \rev{and the propagation of classification errors into downstream scientometric indicators} (Figure~\ref{fig:methodology}). We evaluate encoder-based models under standard fine-tuning and long-tail-aware training settings, and compare them against instruction-tuned LLMs under multiple prompting strategies, including retrieval-augmented prompting and parameter-efficient LoRA adaptation. Finally, we investigate uncertainty-based hybrid encoder--LLM routing to assess whether selective LLM use improves performance on difficult cases without degrading aggregate accuracy.

\subsection*{Dataset}
\label{sec:dataset}

Experiments are conducted on the \textbf{USPTO-70k} benchmark dataset~\citep{pujari2021multitask}, which contains \np{70250} USPTO patents annotated with CPC subclass labels. We focus on subclass-level multi-label classification, a standard benchmark setting in large-scale patent classification. The dataset is chronologically split into \np{50250} training patents (2006--2017), \np{10000} validation patents (2018), and \np{10000} test patents (2019), reflecting realistic deployment conditions.

Each patent includes a title, abstract, and one or more CPC subclass labels. Following standard practice, we use the concatenation of title and abstract as model input. CPC labels are normalized to the four-character subclass level. All models are trained and evaluated using identical textual inputs to ensure comparability.

\subsection*{Supervised encoder baselines}
\label{sec:encoders}

We evaluate \rev{four} supervised Transformer encoder models as strong non-LLM baselines: \textbf{BERT}~\citep{devlin2019bert}, \textbf{SciBERT}~\citep{beltagy2019scibert}, \textbf{PatentSBERTa}~\citep{Bekamiri2024PatentSBERTa}, \rev{and \textbf{BERT-for-Patents}~\citep{srebrovic2020bertpatents}}. The models differ primarily in their pre-training corpora \rev{and objectives}, enabling us to assess the impact of domain-specific pre-training on CPC subclass assignment.

All encoders are fine-tuned as multi-label classifiers using independent sigmoid outputs and binary cross-entropy (BCE) loss. Predictions are obtained by thresholding calibrated probabilities, with the number of predicted subclasses capped at seven per patent to match the LLM evaluation protocol. We additionally evaluate a constrained inference variant in which predictions are restricted to a per-patent subset of candidate subclasses retrieved from CPC definitions. Additional implementation details are reported in Appendix~\ref{app:encoders}.

\subsection*{Long-tail mitigation of encoder models}

To assess whether encoder weaknesses on rare subclasses arise from architectural
limitations or from standard optimisation under class imbalance, we extend \rev{BERT-for-Patents --- the strongest encoder on the validation set ---} with
long-tail-aware training strategies. Specifically, we compare standard binary cross-entropy with focal loss \citep{lin2017focalloss} and asymmetric loss, and evaluate these objectives both with and without oversampling. All configurations use the same encoder architecture, optimisation pipeline, and validation-based threshold calibration procedure as the standard \rev{BERT-for-Patents} baseline. Model selection is based on validation Macro-F1, so that improvements are assessed across the label space, instead of being primarily driven by frequent subclasses (as it is the case for other metrics, such as Micro-F1). Full hyperparameter settings and
results for each configuration are reported in Appendix~\ref{app:longtail}.

\subsection*{Large Language Models}
\label{sec:llms}

We evaluate \rev{five} instruction-tuned LLMs: \textbf{Phi-3 Mini} \citep{Abdin2024Phi3}, \textbf{Mistral-7B} \citep{Jiang2023Mistral}, \textbf{LLaMA-3.1-8B} \citep{Dubey2024LLaMA3}, \rev{\textbf{Qwen3.5-9B} \citep{qwen2026qwen35}, and \textbf{Gemma4-12B} \citep{gemmateam2026gemma4}}. The models span 4-12 billion parameters and are evaluated locally using open weights, in order to preserve reproducibility.
Each LLM is evaluated under four prompting configurations defined by the presence or absence of in-context examples and retrieval-augmented generation (RAG) over CPC subclass definitions (Table~\ref{tab:prompting_configs}).

\begin{table}[h]
\centering
\caption{Prompting configurations defined by in-context examples and CPC definition retrieval.}
\label{tab:prompting_configs}
\small
\begin{tabular}{lcc}
\toprule
 & \textbf{No CPC retrieval} & \textbf{CPC retrieval} \\
\midrule
\textbf{No examples}  & Zero-shot        & Zero-shot + RAG \\
\textbf{Few examples} & Few-shot         & Few-shot + RAG \\
\bottomrule
\end{tabular}

\end{table}

In few-shot settings, prompts include both static and dynamically retrieved examples selected from the training set by semantic similarity. In RAG settings, the LLM receives a restricted set of candidate CPC subclasses together with their official CPC definitions, retrieved using dense vector similarity over CPC definition embeddings. Full prompts and implementation details are provided in Appendices \ref{app:prompts} and \ref{app:rag_details}.

\subsection*{Model Selection and LoRA Fine-Tuning}
\label{sec:lora}

\rev{Qwen3.5-9B achieves the strongest validation performance among the evaluated LLMs and is therefore selected} for parameter-efficient adaptation using LoRA (Low-Rank Adaptation)~\citep{hu2022lora}. LoRA fine-tunes a small number of low-rank adapter parameters while keeping the base-model weights frozen, substantially reducing the number of trainable parameters relative to full fine-tuning.

\rev{To identify the optimal hyperparameter configuration, we conduct a search over LoRA rank, scaling factor, learning rate, and maximum sequence length. The evaluated configurations use ranks $r \in \{16,32\}$, scaling factors $\alpha \in \{32,64\}$, learning rates in $\{1 \times 10^{-4}, 2 \times 10^{-4}\}$, and maximum sequence lengths in $\{512,1024,2048\}$ tokens. Hyperparameter configurations are ranked by validation Micro-F1. Within each run, the checkpoint with the lowest validation loss was retained; the final configuration was then selected using its validation Micro-F1.} Full implementation details \rev{and sweep results} are reported in Appendix~\ref{app:finetuning}.

\subsection*{Hybrid encoder--LLM routing}
\label{sec:hybrid_routing}

Encoders are efficient and accurate on most patents but systematically struggle with ambiguous language, underrepresented subclasses, and cross-domain inventions. We therefore evaluate a selective hybrid pipeline in which an LLM is applied only to patents for which the encoder is uncertain.

The pipeline combines the best-performing adapted encoder \rev{(BERT-for-Patents)} with the fine-tuned \rev{Qwen3.5-9B} model. \rev{BERT-for-Patents} first predicts CPC subclasses for all patents and assigns an uncertainty score to each prediction. Patents are ranked by encoder uncertainty; in particular, we evaluate routing fractions of 2\%, 5\%, 10\%, and 20\%.

In order to reduce the LLM search space and preserve consistency between models, the LLM is restricted to selecting among subclasses identified as plausible either by the encoder itself or through a retrieval-based candidate generation step. Final predictions are obtained through a replacement rule (selected on validation): for routed patents, a valid LLM output replaces the encoder prediction, while non-routed patents retain the encoder output. Candidate selection and replacement procedure are described in Appendix~\ref{app:hybrid}.

\subsection*{Evaluation Metrics and Efficiency Analysis}
\label{sec:metrics}

Models are evaluated using standard metrics for hierarchical multi-label classification~\citep{TsoumakasKatakis2007,SillaFreitas2011}. For each patent, let $Y_i$ denote the set of ground-truth CPC subclasses and $\hat{Y}_i$ the predicted label set. We report micro- and macro-averaged precision, recall, and F1 scores at the CPC subclass level. Micro-averaged metrics aggregate all predictions across the dataset and are therefore dominated by frequent subclasses, while macro-averaged metrics compute scores independently for each subclass and then average across labels, assigning equal weight to frequent and rare categories.

To account for the hierarchical structure of the CPC taxonomy, we additionally report hierarchy-aware precision, recall, and F1 scores computed by augmenting each label with its ancestors in the CPC hierarchy~\citep{SillaFreitas2011}.

Across all experiments, the number of predicted CPC subclasses per patent is capped at seven, reflecting the empirical annotation structure of the dataset and preventing degenerate over-prediction. All models are evaluated under deterministic inference settings.

Aggregate metrics are accompanied by non-parametric bootstrap confidence intervals based on \np{1000} resamples~\citep{Efron1993Bootstrap}. Stratified comparisons between models are evaluated using paired Wilcoxon signed-rank tests over per-label F1 scores~\citep{Hollander2013Nonparametric}.

Moreover, to complement predictive performance with sustainability considerations, we estimate energy usage and associated CO$_2$ emissions using CodeCarbon~\citep{courty_codecarbon_2024}. Measurements are collected for encoder training, LLM inference under different prompting strategies, and LoRA fine-tuning, enabling direct comparison of accuracy--efficiency trade-offs across model families.

Experiments are conducted on the Booster supercomputing system at CINECA using NVIDIA A100 GPUs.

\subsection*{\rev{Classification Bias and Downstream Propagation}}
\label{sec:bias_propagation}

\rev{To assess whether classification errors affect  scientometric conclusions, we compare predictive performance across CPC sections and examine how predicted labels alter aggregated technology indicators. Particular attention is given to Section Y, whose cross-cutting subclasses are widely used to
identify climate- and sustainability-related technologies \citep{veefkind2012epo_ccmt, favot2023green_codes}.}

\rev{For each model, we compare predicted and true test-set patent counts for every CPC subclass. Count distortion is measured using the weighted count error, defined as the total absolute subclass-level count deviation divided by the total number of gold-standard subclass assignments. This formulation prevents subclasses that are extremely rare from disproportionately determining the aggregate error.
We then construct country and assignee rankings separately for each technology, based on assigned patent counts. Ranking distortion is measured as the mean
absolute change in rank position between rankings derived from predicted and true test-set labels. These indicators are calculated for both the complete
taxonomy and Section Y.}

\rev{Then we measure subclass semantic breadth from patent-text embeddings. For each subclass, breadth is the mean cosine distance between each patent embedding and the centroid of the remaining patents assigned
to that subclass; higher values therefore indicate more internally heterogeneous categories. We report Spearman correlations between subclass-level F1, semantic
breadth, training frequency, and test frequency, and regress F1 on semantic breadth, log training frequency, log test frequency, and CPC-section indicators.
False-discovery-rate adjustments are applied across the resulting statistical tests.} Full correlation and regression results are reported in Appendix \ref{app:semantic_breadth}.

\subsection*{External Dataset Construction and Evaluation Protocol (EPO)}

To evaluate out-of-domain generalization, we apply the same experimental pipeline to an external dataset of \np{10000} EPO patent applications filed between 2019 and 2022, obtained from \textit{ORBIS Intellectual Property}.\footnote{\url{https://www.moodys.com/web/en/us/capabilities/company-reference-data/orbis.html}} Patent texts are processed using the same preprocessing and CPC normalization pipeline adopted for the USPTO dataset.

No model is retrained on EPO data. Encoder-based classifiers trained on USPTO patents are evaluated directly in an out-of-domain setting, while LLMs are applied using the same prompting configurations used in the main experiments. Evaluation uses the same flat and hierarchy-aware metrics, frequency stratifications, and efficiency measurements as in the USPTO experiments.

\section*{Results}\label{results}

\subsection*{Data Description}
\label{subsec:data_description}

The patent classification task considered in this study is characterized by a highly imbalanced and multi-label output space. Each patent is annotated with one or more Cooperative Patent Classification (CPC) subclasses at the four-character level (e.g., \texttt{G06F}, \texttt{H04L}), resulting in a large and sparse label vocabulary.

\begin{figure}[t]
    \centering

    \begin{subfigure}[t]{0.98\linewidth}
        \centering
        \includegraphics[width=\linewidth]{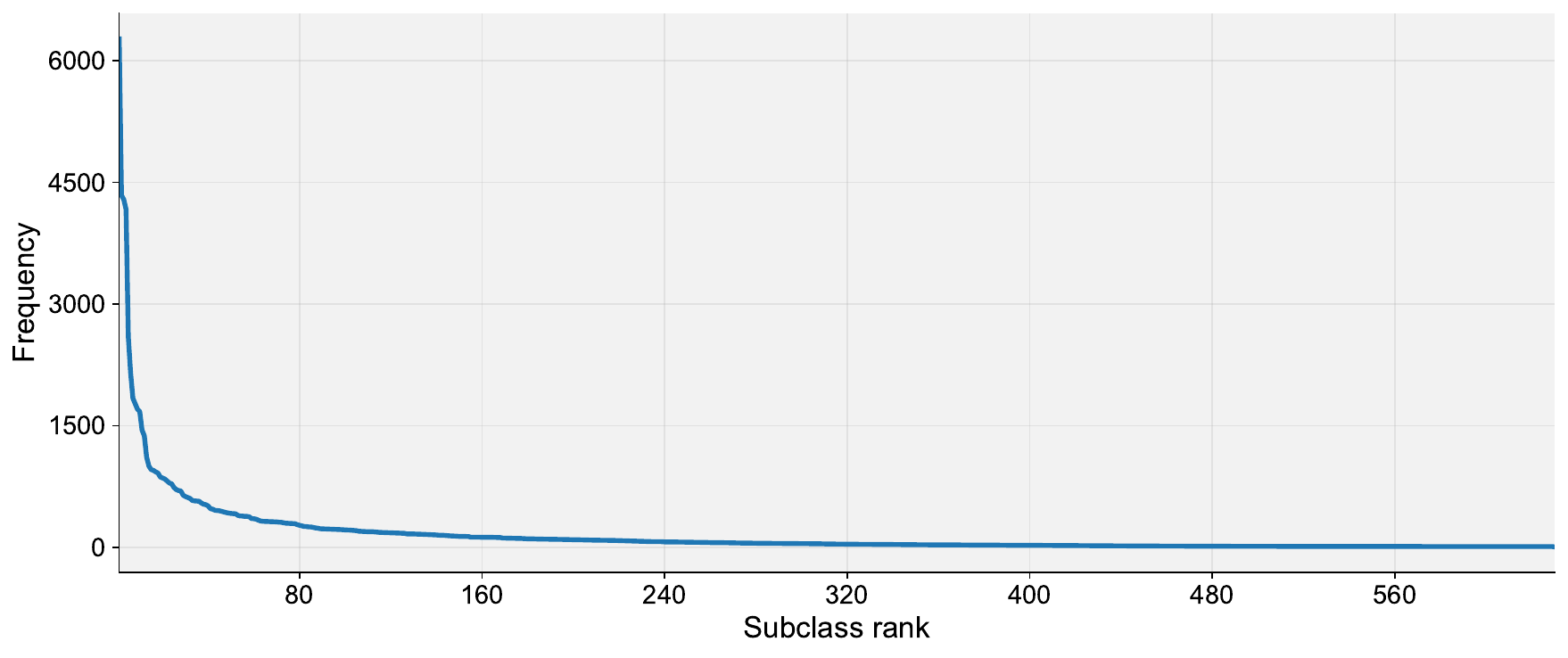}
        \caption{Frequency of CPC subclasses in the training set, ordered by decreasing frequency.}
        \label{fig:subclass_distribution}
    \end{subfigure}

    \vspace{0.6em}

    \begin{subfigure}[t]{0.48\linewidth}
        \centering
        \includegraphics[width=\linewidth]{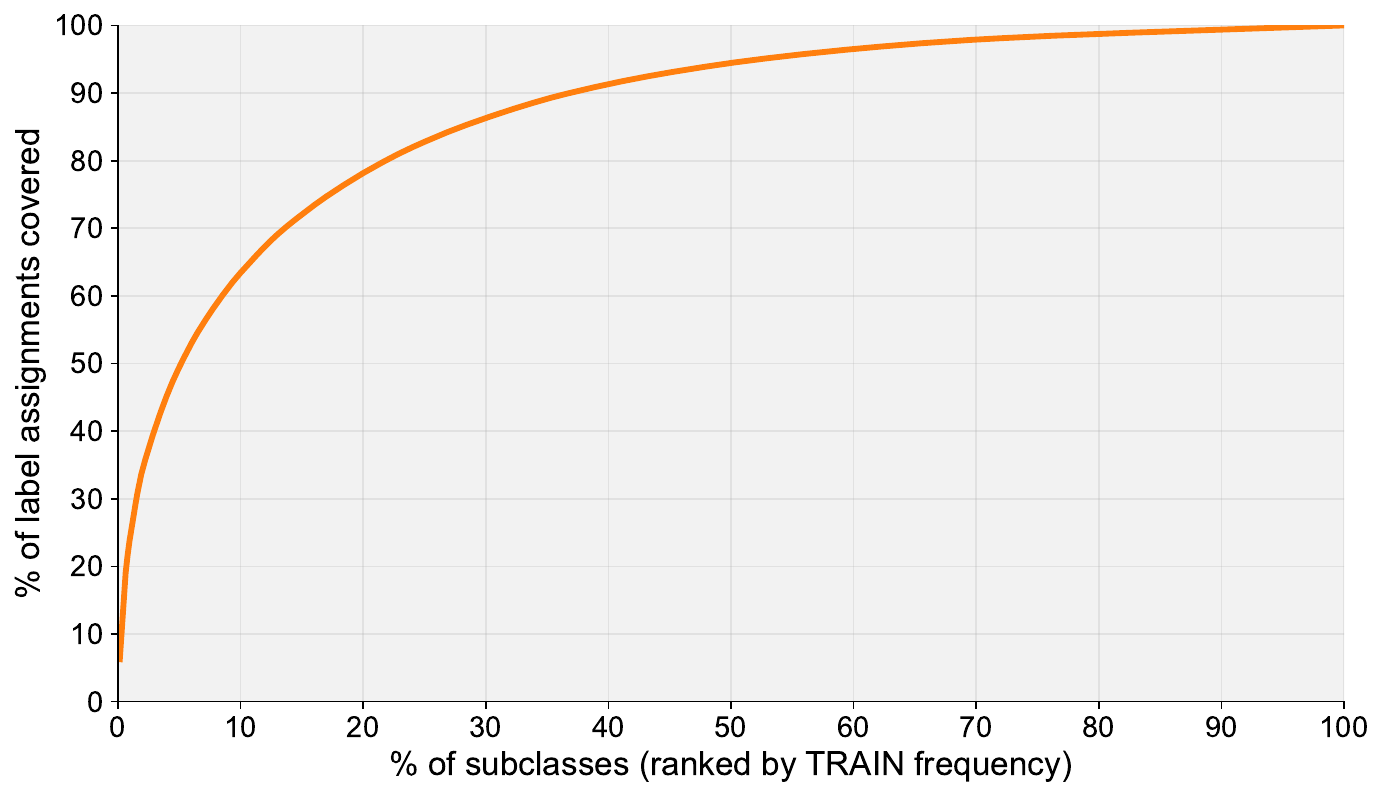}
        \caption{Cumulative coverage of label assignments as a function of the fraction of subclasses.}
        \label{fig:cumulative_coverage}
    \end{subfigure}
    \hfill
    \begin{subfigure}[t]{0.48\linewidth}
        \centering
        \includegraphics[width=\linewidth]{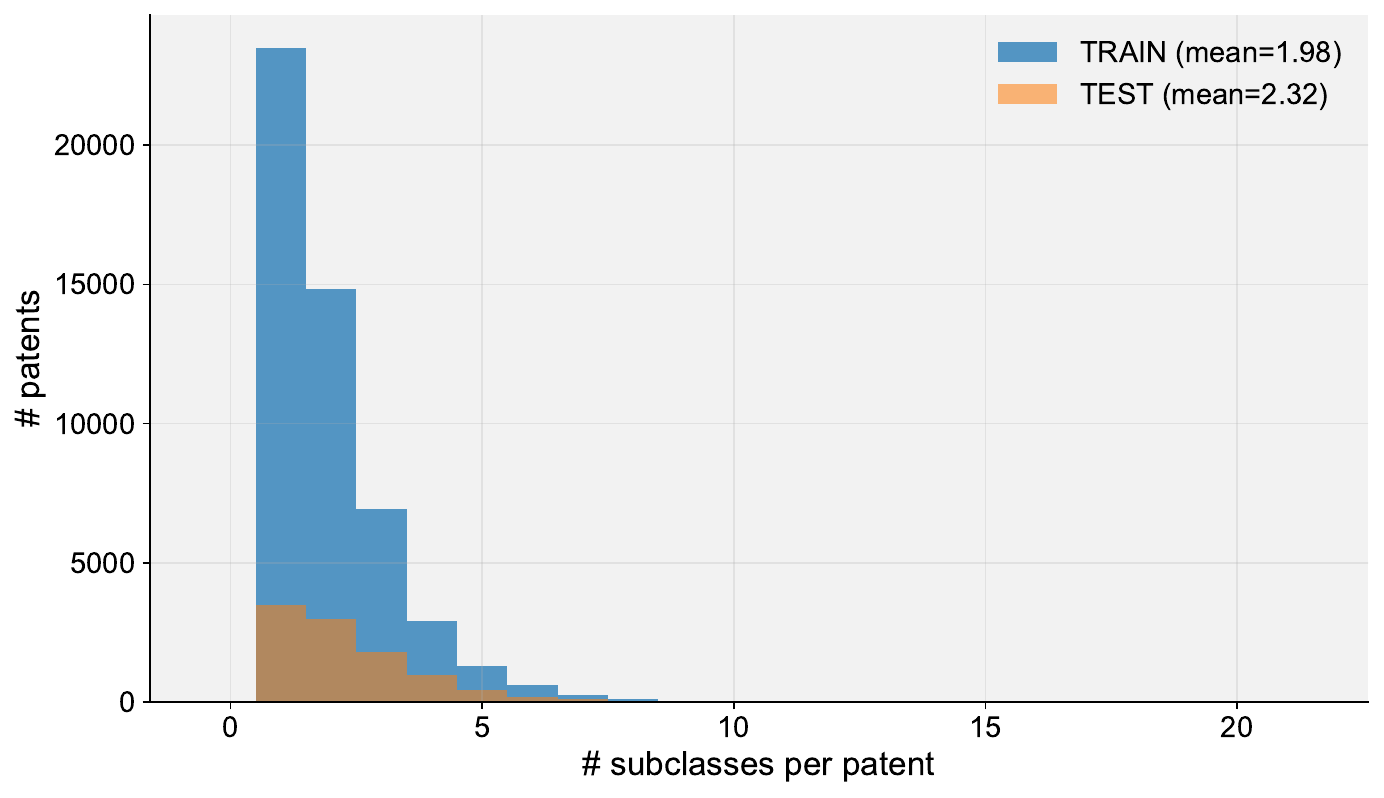}
        \caption{Distribution of the number of CPC subclasses per patent in the training and test sets.}
        \label{fig:labels_per_patent}
    \end{subfigure}

    \caption{
    \textbf{Descriptive statistics of the CPC-labeled patent dataset.}
    (a) Subclass frequency distribution exhibits a pronounced long-tail structure.
    (b) A small fraction of subclasses accounts for the majority of label assignments.
    (c) The distribution of labels per patent is similar across training and test sets, indicating no major shift in label density.
    }
    \label{fig:data_description}
\end{figure}

Figure~\ref{fig:subclass_distribution} reports the distribution of CPC subclasses in the training set, ordered by decreasing frequency. The distribution exhibits a pronounced long-tail structure: a small number of subclasses account for a substantial fraction of all label assignments, while the majority of subclasses appear only rarely. This imbalance is further quantified in Figure~\ref{fig:cumulative_coverage}, which shows that a limited fraction of subclasses covers most of the labeled instances.

In addition, patents are inherently multi-labeled. Figure~\ref{fig:labels_per_patent} reports the distribution of the number of CPC subclasses per patent in both the training and test sets. The two splits exhibit very similar distributions, suggesting the absence of major label-density shifts between training and evaluation data. Further details on the descriptive statistics of the CPC labels in the test and training sets are provided in Appendix \ref{app:cpc_statistics}.

The extreme skewness of the label distribution has important implications for model evaluation. In particular, performance metrics aggregated over all labels (e.g., micro F1) are dominated by a small set of frequent subclasses, while performance on rare and emerging categories contributes little to the aggregate score. This motivates the use of complementary evaluation perspectives and provides context for the comparison between supervised encoder-based models and large language models, which rely less directly on label frequency during inference.


\subsection*{Patent Classification with Off-the-shelf Encoders and LLMs}

We first compare supervised encoder baselines with off-the-shelf LLMs before task-specific adaptation. Table~\ref{tab:main_performance} reports Micro-F1, Macro-F1, and hierarchical Micro-F1 (H-F1) for the four encoders and five LLMs under their best prompting configuration; complete results are provided in Appendix~\ref{app:additional_results}.

Among the encoders, BERT-for-Patents performs best, reaching 0.592 Micro-F1, 0.208 Macro-F1, and 0.673 H-F1. It substantially outperforms BERT, SciBERT, and PatentSBERTa, confirming the value of patent-domain pretraining.

The strongest off-the-shelf LLM is Qwen3.5-9B with few-shot prompting, which achieves 0.525 Micro-F1, 0.320 Macro-F1, and 0.635 H-F1. Gemma4-12B also performs well in the few-shot setting, reaching 0.464, 0.252, and 0.601, respectively. Both models clearly improve on the smaller LLMs, although neither surpasses BERT-for-Patents in Micro-F1 or H-F1. Qwen3.5-9B does, however, obtain a higher Macro-F1 than the standard encoder baseline, indicating more balanced performance across more and less frequent subclasses.

Prompting effects vary across models. Few-shot prompting produces the strongest overall results for Qwen3.5-9B and Gemma4-12B, whereas RAG generally improves Macro-F1 more than aggregate performance (Table \ref{tab:appendix_full_performance}). Overall, recent LLMs are increasingly competitive, but patent-specific pretraining remains more effective before task-specific adaptation.

This initial comparison establishes the strongest off-the-shelf model in each family, but it does not determine whether the remaining gap reflects architectural differences or unequal adaptation to the long-tailed classification task. We therefore next examine whether targeted adaptation strategies strengthen the best-performing encoder (BERT-for-Patents), and whether fine-tuning narrows the gap for the best LLM (Qwen3.5-9B).

\begin{table}[t]
\centering
\small
\setlength{\tabcolsep}{5pt}
\caption{
Main CPC subclass classification results on the USPTO test set. The table reports
the strongest off-the-shelf configuration for each principal model family, the
task-adapted encoder and LLM, and the selected hybrid configuration. Complete
results for all encoder baselines, LLMs, prompting strategies, long-tail training
configurations, and routing fractions are reported in Appendix \ref{tab:appendix_full_performance}.
\label{tab:main_performance}
}
\begin{tabular}{lccc}
\toprule
Model / configuration & Micro-F1 & Macro-F1 & H-F1 \\
\midrule
\multicolumn{4}{l}{\textit{Best off-the-shelf configurations}} \\
BERT-for-Patents & 0.592 & 0.208 & 0.673 \\
Qwen3.5-9B (few-shot) & 0.525 & 0.320 & 0.635 \\
Gemma4-12B (few-shot) & 0.464 & 0.252 & 0.601 \\
\midrule
\multicolumn{4}{l}{\textit{Task-adapted configurations}} \\
BERT-for-Patents (long-tail) &
\textbf{0.616} & 0.410 & 0.694 \\
Qwen3.5-9B + LoRA &
0.602 & 0.379 & 0.686 \\
\midrule
\multicolumn{4}{l}{\textit{Hybrid configuration}} \\
Hybrid 5\% &
0.614 & \textbf{0.412} & \textbf{0.695} \\
\bottomrule
\end{tabular}
\end{table}

\subsection*{Long-tail-aware Encoder Training and LoRA Adaptation}

In terms of long-tail adaptation strategies, for BERT-for-Patents, we compare binary cross-entropy, focal loss, and asymmetric loss, with the latter objectives also combined with oversampling. The strongest validation configuration uses asymmetric loss with oversampling (Appendix \ref{app:encoders}).

On the test set, the long-tail-aware encoder reaches 0.616 Micro-F1, 0.410 Macro-F1, and 0.694 H-F1, compared with 0.592, 0.208, and 0.673 for the standard BERT-for-Patents baseline. The largest gain is, therefore, observed in Macro-F1, showing that imbalance-aware training substantially improves performance across less frequent subclasses.

Qwen3.5-9B is adapted through a targeted LoRA sweep over rank, scaling factor, learning rate, and maximum sequence length. The selected configuration uses $r=16$, $\alpha=32$, a learning rate of $1\times10^{-4}$, and a maximum sequence length of 1024 tokens. It reaches 0.602 Micro-F1, 0.379 Macro-F1, and 0.686 H-F1, compared with 0.525, 0.320, and 0.635 for the strongest off-the-shelf configuration.

Therefore, both model families benefit substantially from task-specific adaptation. Long-tail-aware BERT-for-Patents remains the strongest overall model, although LoRA adaptation considerably narrows the performance gap. Aggregate metrics alone, however, do not establish whether the capabilities of the two model families (encoders and LLMs) are alternative or complementary. Therefore, we next test whether the LLM can improve the specific patents on which the encoder is least reliable.

\begin{figure*}[t]
    \centering

    \begin{subfigure}[t]{0.85\linewidth}
        \centering
        \includegraphics[width=\linewidth]{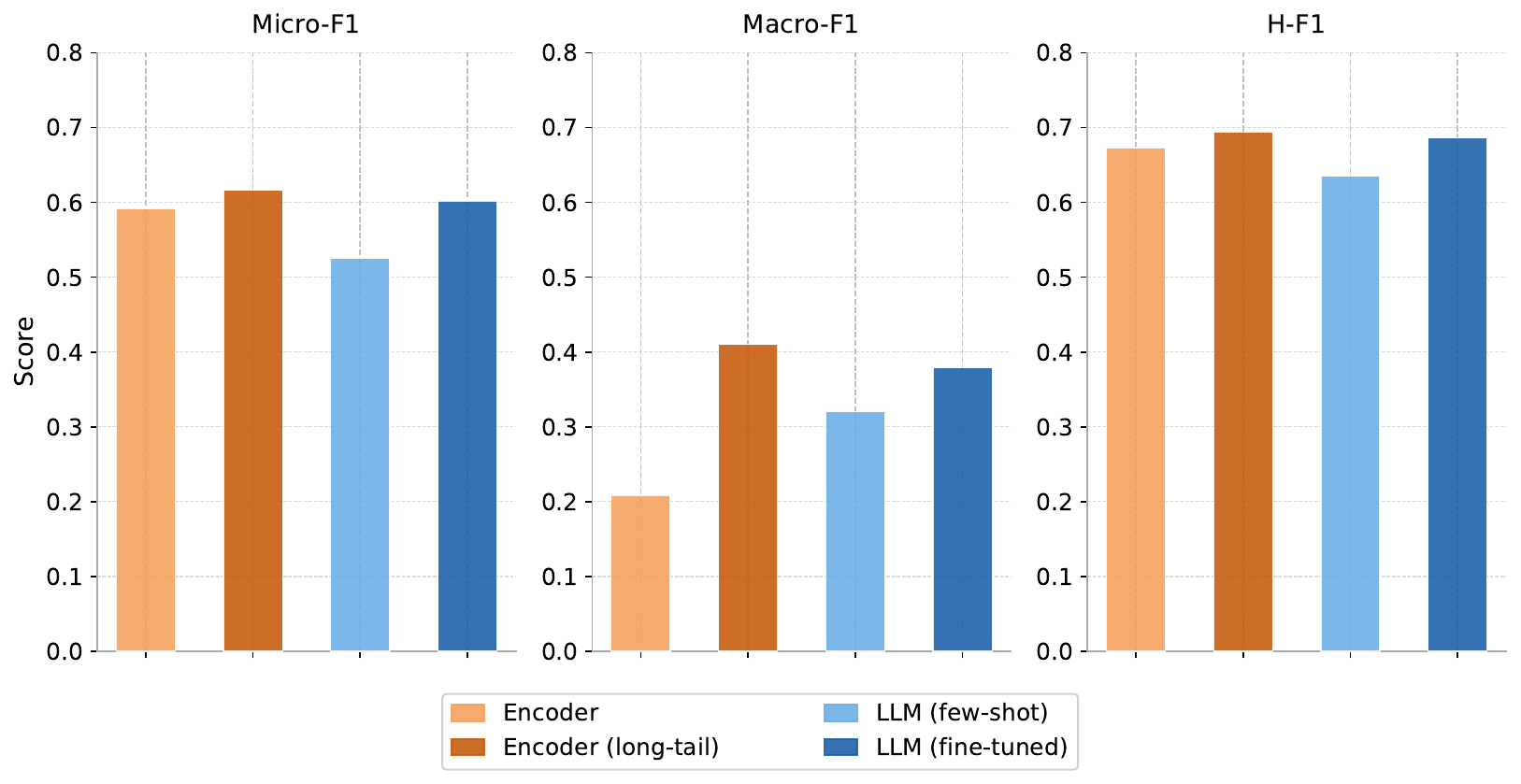}
        \caption{
        Long-tail-aware training substantially improves BERT-for-Patents across aggregate, macro-level, and hierarchy-aware metrics, while LoRA adaptation considerably improves Qwen3.5-9B relative to few-shot prompting.
        }
        \label{fig:longtail_lora_comparison}
    \end{subfigure}
    \hfill
    \begin{subfigure}[t]{0.85\linewidth}
        \centering
        \includegraphics[width=\linewidth]{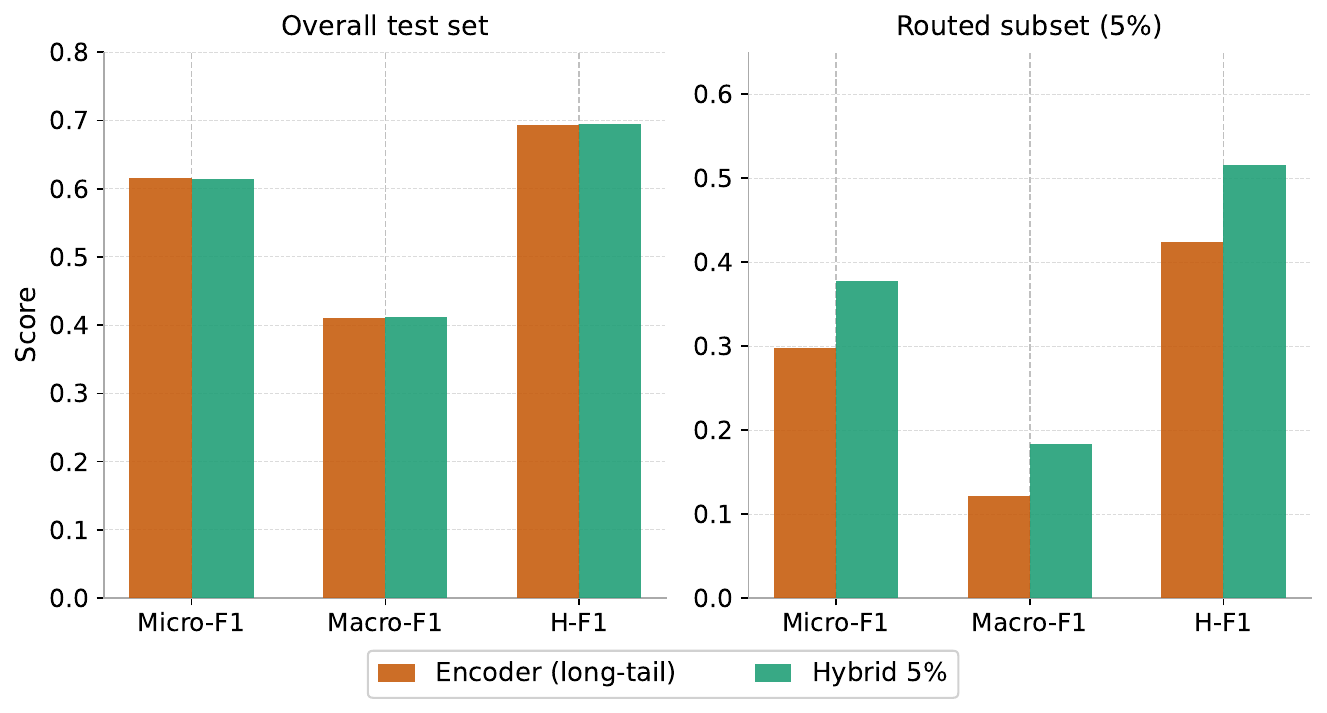}
        \caption{
        Hybrid routing leaves full-test-set performance nearly unchanged, but improves performance on the routed subset of uncertain patents, particularly on hierarchy-aware metrics.
        }
        \label{fig:hybrid_routing}
    \end{subfigure}

    \caption{
    Effects of task-specific adaptation (a) and selective hybrid routing (b) on classification performance at subclass level.
    }
    \label{fig:adaptation_hybrid}
\end{figure*}

\subsection*{Hybrid Encoder--LLM Routing}

We evaluate this complementarity through a hybrid pipeline based on encoder uncertainty, combining long-tail-aware BERT-for-Patents with Qwen3.5-9B-LoRA. The encoder first classifies all patents, after which the 2\%, 5\%, 10\%, or 20\% most uncertain cases are routed to the LLM.

At the full-test-set level, routing produces only small changes. The 5\% hybrid provides the best balance, with 0.614 Micro-F1, 0.412 Macro-F1, and 0.695 H-F1, whereas broader routing does not improve aggregate performance.
Nevertheless, the gains are much larger on the routed subsets. Micro-F1 increases from 0.210 to 0.338 at 2\% routing, from 0.298 to 0.378 at 5\%, and from 0.377 to 0.418 at 10\%. At 20\%, the improvement is negligible, from 0.448 to 0.451.

Paired Wilcoxon tests over per-subclass F1 scores confirm significant improvements at 2\%, 5\%, and 10\% routing after Bonferroni correction, whereas the 20\% configuration is not significant. Effect sizes also decline as the routed fraction increases.

The hybrid therefore provides complementary value primarily for a small subset of highly uncertain patents and is best interpreted as a targeted refinement mechanism, rather than as a general replacement for the encoder. However, neither aggregate performance nor routed-subset gains reveal where the remaining errors occur or whether they matter for downstream analysis. We therefore turn from model-level performance to the distribution and consequences of classification errors across the CPC taxonomy.

\begin{figure*}
    \centering

    \begin{subfigure}[t]{1.0\textwidth}
        \centering
        \includegraphics[width=0.7\linewidth]{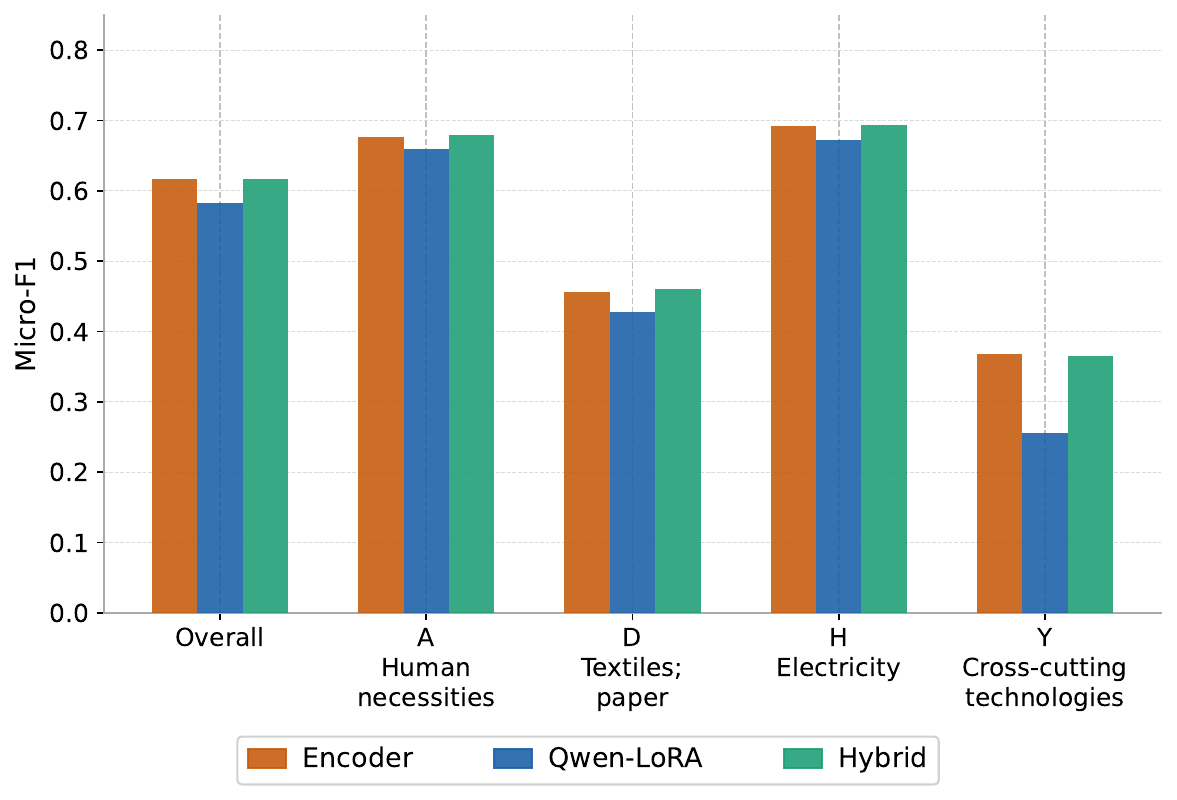}
        \caption{
        Subclass-level Micro-F1 for the complete test set and selected CPC
        sections. Section Y is substantially more difficult than the complete
        taxonomy and conventional CPC sections. Hybrid routing does not
        recover this gap.
        }
        \label{fig:bias_propagation_a}
    \end{subfigure}

    \vspace{0.6em}

    \begin{subfigure}[t]{0.49\textwidth}
        \centering
        \includegraphics[width=0.98\linewidth]{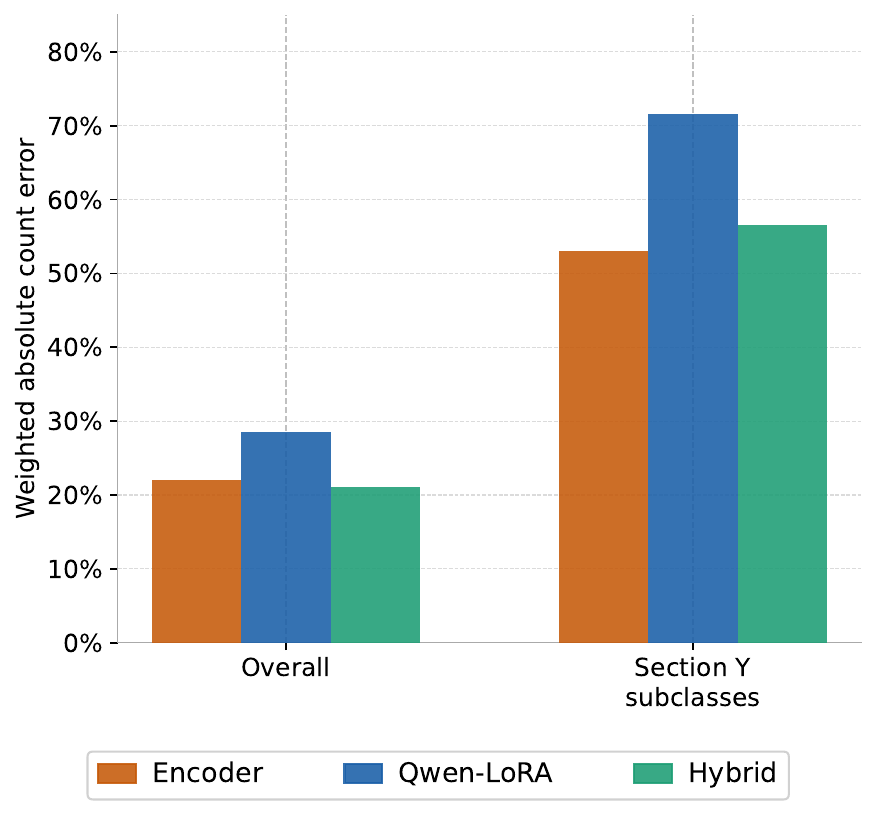}
        \caption{
        Support-weighted absolute count error across all CPC subclasses and Section Y subclasses. The metric aggregates absolute differences between predicted and gold-standard subclass counts and normalises them by the total number of gold-standard assignments. All models produce substantially larger count distortions within Section Y.
        }
        \label{fig:bias_propagation_b}
    \end{subfigure}
    \hfill
    \begin{subfigure}[t]{0.49\textwidth}
        \centering
        \includegraphics[width=1.\linewidth]{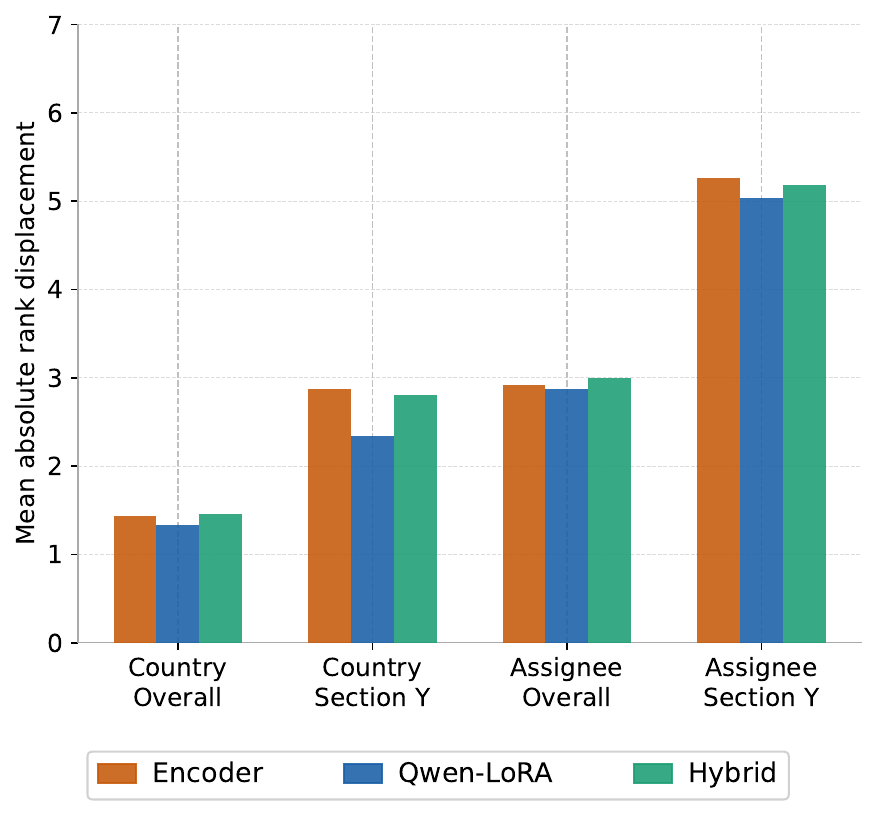}
        \caption{
        Mean absolute displacement of country and assignee positions in
        technology rankings constructed from predicted rather than
        gold-standard CPC assignments. Ranking distortion is substantially
        larger for Section Y, especially at the assignee level.
        }
        \label{fig:bias_propagation_c}
    \end{subfigure}

    \caption{
    Classification bias and its propagation into downstream scientometric
    indicators. Panel~\subref{fig:bias_propagation_a} shows section-specific
    predictive performance, Panel~\subref{fig:bias_propagation_b} shows
    distortion in the estimated technological composition, and
    Panel~\subref{fig:bias_propagation_c} shows the resulting displacement in
    country and assignee rankings.
    }
    \label{fig:bias_propagation}
\end{figure*}

\rev{\subsection*{From Classification Errors to Downstream Measurement Distortion}
\label{sec:bias_propagation}

The previous analyses establish which models perform best overall, but not whether their remaining errors are concentrated in particular technological domains or
whether models with similar F1 scores produce equally reliable scientometric indicators. This is a particularly important question because classification choices can shape the measurement and interpretation of inventive activity, even when aggregate predictive performance appears comparable \citep{Lobo2019InventiveNoveltyClassification}. In this section, we examine classification difficulty across CPC categories and its propagation into technology counts and rankings of countries and assignees.

Figure~\ref{fig:bias_propagation} summarises the performance comparison at the level of CPC sections. The encoder achieves the strongest standalone performance overall, with a Micro-F1 of 0.616, compared with 0.602 for Qwen3.5-LoRA and 0.614 for the 5\% hybrid. Performance is, however, substantially lower in Section Y: Micro-F1 falls to 0.443 for the encoder,
0.393 for Qwen3.5-LoRA, and 0.439 for the hybrid.

Is this weakness driven by label frequency/rarity, or by semantic heterogeneity within sections? We measure semantic breadth as the mean cosine distance between the patents assigned to a subclass and their subclass centroid. We find that across all subclasses, breadth is negatively correlated with F1:
Spearman's $\rho$ is $-0.257$ for the encoder, $-0.235$ for Qwen3.5-LoRA, and
$-0.280$ for the hybrid. Instead, training frequency is positively associated with F1, with correlations of 0.239, 0.287, and 0.247, respectively. All associations remain significant after FDR correction.

Multivariate regressions confirm that semantic breadth is negatively associated with F1 after controlling for log training frequency, log test support, and CPC
section. The estimated coefficients are $-6.076$ for the encoder, $-5.845$ for Qwen3.5-LoRA, and $-6.447$ for the hybrid, with FDR-adjusted $p<0.001$ in every case. Classification difficulty therefore depends not only on the number of available examples, but also on the internal heterogeneity and boundary ambiguity of the
categories being predicted. This helps explain the persistent difficulty of Section Y, whose cross-cutting categories overlap with several conventional technological domains \citep{veefkind2012epo_ccmt,favot2023green_codes,rainville2025circular_patents}. We report full correlation and regression results in Appendix \ref{app:semantic_breadth}.

We then examine count distortion by comparing, for each subclass, the number of patents obtained from predicted and true test-set labels. To prevent rare subclasses from disproportionately determining the aggregate result, we report the weighted absolute count error, computed as the sum of the absolute subclass-level count deviations divided by the total number of gold-standard subclass assignments. For the complete taxonomy, the weighted error is 21.98\% for the encoder, 28.47\% for Qwen3.5-LoRA, and 21.08\% for the 5\% hybrid. Within Section Y, it increases sharply to 53.04\%, 71.57\%, and 56.58\%, respectively. Thus, all models produce substantially greater distortions in estimated technological prevalence within Section Y. The hybrid yields the lowest count distortion across the complete taxonomy, whereas the encoder is more reliable within Section Y.

Finally, we rank countries and assignees within each technology according to their patent counts and compute the mean absolute displacement from the corresponding
gold-standard rankings. Across the full taxonomy, country-rank displacement is 1.43 positions for the encoder, 1.34 for Qwen3.5-LoRA, and 1.46 for the hybrid;
within Section Y, it rises to 2.87, 2.34, and 2.81. Assignee-rank displacement similarly increases from 2.91, 2.87, and 3.00 positions overall to 5.27, 5.03,
and 5.18 within Section Y.

Predictive accuracy and downstream validity are therefore related but not perfectly aligned. The 5\% hybrid produces the lowest weighted count error across the complete taxonomy, whereas the encoder is the most reliable model for Section Y counts.
Ranking results reveal a different ordering. Qwen3.5-LoRA has the lowest Micro-F1 of the three models, but it produces the smallest country- and assignee-rank displacement both overall and within Section Y, although the differences are generally modest. Model choice should therefore depend on the intended application: the encoder or hybrid may be preferable when accurate patent-level classification and technology counts are the primary objective, whereas LLMs may preserve relative actor positions slightly better under the present ranking metric. Model validation should consequently consider not only Micro-F1 and Macro-F1, but also
the reliability of the specific counts and rankings that the predictions are intended to support.}

\subsection*{Energy--Accuracy Trade-off}

Another important dimension in practice is the computational cost of deploying patent-classification models at scale. The preceding results show that the long-tail-aware encoder provides the strongest overall predictive performance, while the LLM adds value mainly when applied selectively to patents on which the encoder is highly uncertain. We therefore examine how these differences translate into inference time, energy consumption, and overall deployment efficiency.

\begin{figure}[t]
    \centering  \includegraphics[width=0.9\linewidth]{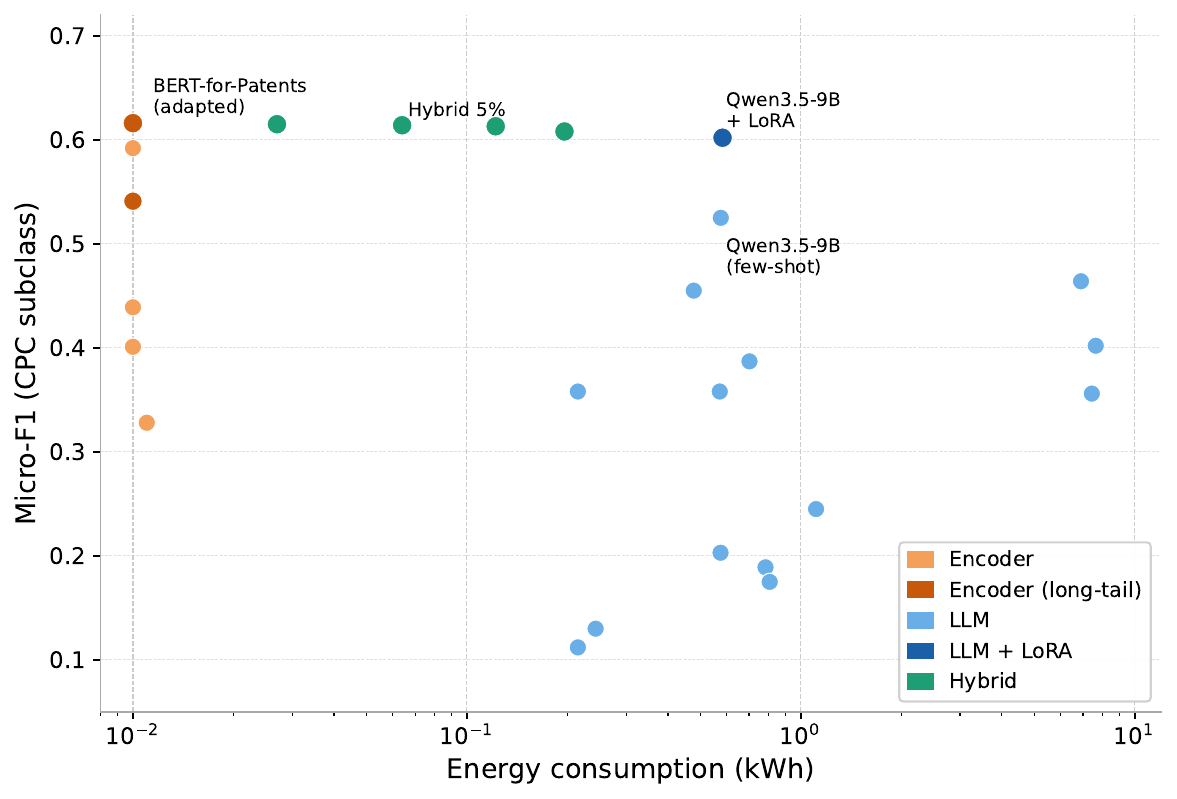}
    \caption{Accuracy–energy trade-off across model families on CPC subclass classification (N=10,000 patents). Energy refers to inference cost measured with CodeCarbon. Encoder models (orange) achieve the best trade-off; the long-tail-aware encoder variant reaches the highest Micro-F1 at minimal energy cost. LLMs (blue) consume one to two orders of magnitude more energy with lower accuracy across all prompting regimes. Qwen-LoRA denotes Qwen3.5-9B fine-tuned with LoRA adaptation. The hybrid system combines BERT-for-Patents (long-tail) with Qwen-LoRA via uncertainty-based routing and is represented in green.} 
    \label{fig:accuracy_energy_tradeoff}
\end{figure}

Figure \ref{fig:accuracy_energy_tradeoff} compares predictive performance and computational cost across the final model configurations; full numerical results are reported in Table \ref{tab:cost_time_energy_co2}. Full-test-set inference with Qwen3.5-9B-LoRA requires approximately 101 minutes for \np{10000} patents, consumes 0.583 kWh, and produces an estimated 0.0023 kg of CO$_2$. This corresponds to approximately 0.606 seconds and 0.058 Wh per patent.

As seen, this higher computational cost is not offset by superior predictive performance.
The hybrid configurations offer a more selective deployment strategy by restricting Qwen inference to the 2-20\% of patents for which the encoder is most uncertain. The 5\% hybrid provides the strongest overall balance, reaching 0.614 Micro-F1, 0.412 Macro-F1, and 0.695 hierarchical F1 while routing only 500 of the \np{10000} test patents. This preserves nearly all of the encoder's aggregate performance while avoiding the cost of applying the LLM to the complete corpus.
Overall, the long-tail-aware encoder remains the most effective and efficient foundation for large-scale patent classification. The LLM is more plausibly deployed as a selective refinement mechanism, where its additional computational cost is incurred only for the limited subset of cases in which it provides measurable gains.

\subsection*{Validation on external data (EPO)}

To assess external validity, we evaluate the final models on an independent set of
\np{10000} EPO patents filed between 2019 and 2022, obtained from ORBIS Intellectual Property. The models are trained on USPTO data and applied to the EPO sample without retraining, introducing both institutional and
temporal domain shift.
The EPO results broadly confirm the main findings
(Appendix~\ref{app:epo_results}). Long-tail-aware BERT-for-Patents remains the strongest standalone model, while LoRA adaptation substantially improves Qwen3.5-9B without allowing it to surpass the encoder. The 5\% hybrid achieves essentially the same Micro-F1 as the adapted encoder, with small gains in Macro-F1
and hierarchy-aware F1. On the routed subset, however, the hybrid produces a substantial and statistically significant improvement over the encoder.
These results indicate that the relative strengths of the encoder, adapted LLM, and selective hybrid are robust to institutional and temporal domain shift.

\section*{Discussion}\label{discussion}

The comparison between encoders and LLMs points to a clear hierarchy of determinants in patent classification. Unsurprisingly, domain-specific pre-training matters most: among off-the-shelf systems, BERT-for-Patents performs better than both generic encoders and prompted LLMs. This suggests that familiarity with patent language and drafting conventions remains more valuable than broad generative capacity when no task-specific adaptation is applied, consistent with broader evidence that fine-tuned encoder models can outperform substantially larger prompted language models in supervised text
classification settings, particularly when sufficient labelled data are available
\citep{edwards-camacho-collados-2024-language}.

Supervised adaptation changes the comparison, but not the ranking. Long-tail-aware BERT-for-Patents achieves the best overall performance, while LoRA fine-tuning substantially improves Qwen3.5-9B and brings it close to the encoder. The implication is twofold. First, evaluations based only on zero-shot or few-shot prompting understate what LLMs can achieve after task-specific training. Second, the larger generative model still does not surpass the smaller patent-specific encoder once both are adapted seriously. For CPC classification, architectural scale alone is therefore a poor predictor of practical value.

The gains from long-tail-aware training are especially important. The large increase in Macro-F1 indicates that asymmetric loss and moderate oversampling improve coverage across the label space rather than merely reinforcing already frequent categories. 
The hybrid experiments reveal, instead, a narrower form of complementarity. Qwen improves predictions for patents on which the encoder is least confident, and these gains are statistically significant at lower routing fractions. Once the routed share expands, however, the advantage weakens and eventually disappears. The practical lesson is that LLM use should be targeted to very specific cases. A hybrid system is useful when it identifies cases with a high probability of encoder failure; simply sending more patents to the LLM adds cost without producing comparable benefits. Importantly, our hybrid approach differs from previous studies, which combine models across the complete classification workflow or use LLMs to support iterative
training-data construction \citep{Kamateri2023Ensemble,xiong2025scalable}.

A second contribution concerns where classification errors occur. Performance is markedly weaker in Section Y, and this pattern persists across all the considered architectures. This difficulty cannot be attributed to label scarcity alone. We find that semantic breadth remains negatively associated with subclass-level F1 after controlling for training frequency, test support, and CPC section. Broad and internally heterogeneous categories are harder to predict because they encompass less cohesive technological content and often overlap with several
conventional domains, a challenge that is consistent with prior work on hierarchical, interdisciplinary, and temporally evolving patent categories
\citep{Xu2025InterdisciplinaryPatents}.

This finding has implications beyond model selection. Section Y contains cross-cutting categories, extensively used in studies of climate and sustainability technologies \citep{veefkind2012epo_ccmt, favot2023green_codes, rainville2025circular_patents}. If these categories are systematically more difficult to classify, then errors are likely to be concentrated precisely in domains of high policy and scientometric relevance. The problem is therefore not only that some labels are rare, but that parts of the taxonomy are intrinsically harder to operationalise because their boundaries are broader and less stable.
The downstream analysis confirms that these errors affect patentometric measurement. Differences in document-level F1 translate into distortions in subclass counts and shifts in country and assignee rankings. Moreover, the ordering of models depends on both the downstream quantity and the technological scope considered: the 5\% hybrid produces the lowest weighted count error across the complete taxonomy, whereas the encoder is more reliable for Section Y counts. Qwen3.5-9B-LoRA produces numerically smaller actor-rank displacement despite weaker predictive performance. Predictive accuracy and measurement validity are thus related, but distinct.

This matters for patent-based research because CPC assignments are routinely aggregated into indicators of technological activity, specialisation, diversification, and leadership. A classifier can perform well on average and still bias conclusions in a specific domain. As also discussed in previous studies, classification-system choices can themselves influence the resulting interpretation
of inventive activity \citep{Lobo2019InventiveNoveltyClassification}. Therefore, validation and model choice should be aligned with the intended use of the predictions. For descriptive mapping, count distortion may be central; for comparative analysis, ranking stability may matter more; for operational classification, document-level precision and recall may remain the primary concern.

Computational efficiency further strengthens the case for the encoder as the default model. Full-corpus inference with Qwen3.5-9B-LoRA requires substantially more time and energy without improving predictive performance. This aspect reinforces broader concerns about the large resource consumption of generative-model
inference and the importance of evaluating accuracy jointly with computational cost
\citep{schwartz2020green,niu2025energy}. The hybrid offers a more credible deployment strategy because it limits autoregressive inference to a small set of uncertain patents.

We acknowledge several limitations, which also point to promising directions for future research. First, the analysis is restricted to CPC subclasses; extending the
evaluation to the finer main-group and subgroup levels may reveal different relative strengths of encoders, LLMs, and hybrid systems. Second, the comparison is
intentionally limited to open-weight LLMs that can be deployed on the same computational infrastructure, enabling a more controlled assessment of predictive
performance, efficiency, and energy consumption.
Third, the mitigation strategies considered here remain predominantly frequency-based. A more domain-specific approach would integrate CPC definitions, notes, references, exclusions, and hierarchical relations directly into model training, allowing models to learn the technical meaning and boundaries of categories. Contrastive alignment between patent texts and enriched CPC representations appears particularly promising for
sparse, newly introduced, or semantically diffuse classes.

More broadly, future work could complement textual and taxonomic information with additional patent metadata and network-based representations. Previous work has shown that patent networks and graph embeddings can enrich classification and landscaping
beyond document text alone
\citep{Liu2011HybridPatent,Choi2022PatentLandscaping}. Patent datasets encode relations among inventors, assignees, technological fields, and collaborative structures
that may provide information not captured by document text alone. Recent work on the structure of innovation networks, and on scalable network embeddings based on
approximate equitable partitions suggests promising ways to represent such structural information efficiently \citep{emer2026hidden,squillace2024efficient,squillace2026scalable}.
Combining text, enriched CPC representations, metadata, and network structure may therefore support more robust classification, particularly for rare and cross-cutting
technological categories.

\section*{Conclusion}
\label{conclusion}

Automated patent classification is increasingly used to support large-scale technological analysis, yet it remains unclear how strong encoder models and LLMs compare once both are properly adapted, where they complement one another, and how their errors affect the indicators built from predicted labels.

Our findings indicate that patent-specific encoders remain the most reliable basis for large-scale CPC classification. Long-tail-aware BERT-for-Patents delivers the strongest overall performance and the best accuracy-efficiency trade-off, while LoRA adaptation makes Qwen3.5-9B substantially more competitive without allowing it to surpass the encoder. The hybrid architecture adds value in a more limited but meaningful role: it identifies a small subset of highly uncertain patents for which the LLM produces significant gains, whereas broader routing offers little additional benefit.
The analysis also reveals that classification difficulty is not driven by label frequency alone. Errors concentrate in semantically broad and cross-cutting categories, especially Section Y, and these errors propagate into distorted technology counts and unstable country and assignee rankings. This is important because models with similar aggregate performance can support different conclusions about technological activity and leadership.
We argue that in practice, patent-specific encoders should remain the default option for large-scale deployment, with adapted LLMs used selectively where they provide a clear comparative advantage. Evaluation should therefore extend beyond Micro-F1 and Macro-F1 to include domain-specific error patterns, indicator stability, and computational cost.

\backmatter

\bmhead{Data and code availability}

The USPTO 70k dataset is publicly available and can be accessed online \citep{pujari2021multitask}. The EPO data obtained through ORBIS IP of Bureau van Dijk are subject to licensing restrictions and cannot be redistributed, but our results are replicable by any researcher with access to ORBIS IP. To support transparency and reproducibility, the core experimental code used in this study is publicly available at
\url{https://github.com/lorenzoemer/Encoder-Based-Models-vs-Large-Language-Models-for-Patent-Classification}.

\bmhead{Acknowledgements}
The authors acknowledge ISCRA for awarding the project \emph{PATTERNS} access to the LEONARDO supercomputer, owned by the EuroHPC Joint Undertaking and hosted by CINECA (Italy), which enabled the experiments reported in this paper.
The work has been partially supported by project SMaRT COnSTRUCT (CUP J53C24001460006), in the context of FAIR (PE0000013, CUP B53C22003630006) under  the National Recovery and Resilience Plan (Mission 4, Component 2, Line of Investment 1.3) funded by the European Union - NextGenerationEU.

\section*{Declarations}
\bmhead{Conflict of interest}
The authors have no relevant financial or non-financial interests to disclose.

\clearpage

\bibliography{sn-bibliography}

\clearpage

\begin{appendices}

\section{Supervised encoder models: implementation details}
\label{app:encoders}

\subsection{Model architecture and training}

We evaluate four encoder architectures: BERT, SciBERT, PatentSBERTa, and
BERT-for-Patents. Each model is fine-tuned for multi-label CPC classification by
attaching a linear classification head that outputs one logit for each CPC
subclass. The logits are transformed into independent probabilities using a
sigmoid function, and the baseline models are trained using binary
cross-entropy over the complete label set.

All models are trained on the same chronological training split. Hyperparameter
selection, input-length selection, threshold calibration, and checkpoint
selection are conducted exclusively on the validation split, while final
performance is evaluated once on the held-out test set. Model inputs consist of
the patent title concatenated with the abstract.

Maximum input length is treated as a validation-tuned hyperparameter rather than
fixed a priori. For the baseline encoder comparison, alternative input lengths
are evaluated on the validation set and the best-performing configuration is
retained for each model. The long-tail-aware BERT-for-Patents configurations are
subjected to a separate validation-based comparison, with the retained sequence
lengths reported in Section~\ref{app:longtail}.

\subsection{Threshold calibration and label decoding}

For each encoder, the prediction threshold is calibrated on the validation set
to maximise micro-averaged F$_1$. At test time, a CPC subclass is predicted when
its calibrated probability exceeds the model-specific threshold. To ensure
comparability with the LLM outputs, we cap the number of predicted subclasses at
seven per patent. If no probability exceeds the calibrated threshold, the
subclass with the highest predicted probability is retained.

CPC subclass assignments are sparse. In the training set
(\np{50250} patents), the mean number of subclasses per patent is 1.98 and the
median is 2; 90\% and 95\% of patents have at most four subclasses, while 99\%
have at most seven. Overall, 99.48\% of training patents are assigned between
one and seven CPC subclasses, with a maximum of 18.

The test set exhibits a similar distribution, with a mean of 2.32 subclasses,
a median of 2, and a 99th percentile of seven. Overall, 99.21\% of test patents
have between one and seven subclasses. We therefore restrict outputs from all
encoder- and LLM-based methods to a maximum of seven subclasses. This constraint
prevents degenerate over-prediction while remaining consistent with the
empirical annotation distribution.

\subsection{Constrained label-space evaluation}

In addition to evaluation over the complete CPC subclass space, we consider a
constrained decoding setting that mirrors the retrieval-based label-space
restriction used for LLM prompting. For each patent, the official CPC subclass
definitions are embedded using the E5-base-v2 bi-encoder and ranked by cosine
similarity to the patent text. The top-$K$ subclasses, with $K=20$, form the
patent-specific candidate set.

During constrained evaluation, probabilities assigned to subclasses outside
the retrieved top-$K$ set are masked before thresholding and label decoding. The
value $K=20$ determines only the size of the candidate label space; the final
number of predicted subclasses remains capped at seven. This restriction is
applied exclusively at inference time and does not affect model training or
parameter estimation.

\subsection{Long-tail mitigation: training details and configuration comparison}
\label{app:longtail}

To evaluate the impact of long-tail-aware training on encoder performance, we
retrain BERT-for-Patents using binary cross-entropy (BCE), focal loss, and
asymmetric loss (ASL), with focal loss and ASL additionally combined with
label-frequency-aware oversampling. All retained configurations use a linear
multi-label classification head.

Maximum input length is validated separately for the baseline and long-tail-aware
training settings. The retained BCE configuration uses a maximum sequence length
of 512 tokens, whereas the retained focal-loss and ASL configurations use 256
tokens. These values are selected on the validation set.

Focal loss~\citep{lin2017focalloss} replaces standard BCE with a modulated
objective that down-weights well-classified examples:
\begin{equation}
  \mathcal{L}_{\mathrm{FL}}
  =
  -\alpha_t(1-p_t)^{\gamma}\log(p_t),
\end{equation}
where $p_t$ is the predicted probability of the correct class, $\gamma$
controls the degree of down-weighting, and $\alpha_t$ is an optional
class-balancing factor.

ASL extends this principle by applying different focusing terms to positive and
negative labels. For label $j$, the loss is
\begin{equation}
  \mathcal{L}_{\mathrm{ASL}}
  =
  -y_j(1-p_j)^{\gamma_{+}}\log(p_j)
  -(1-y_j)(p_j^{\,m})^{\gamma_{-}}\log(1-p_j^{\,m}),
\end{equation}
where $\gamma_{+}$ and $\gamma_{-}$ independently control the contributions of
positive and negative examples, and $p_j^{\,m}$ denotes the optionally clipped
negative probability. By suppressing abundant easy negatives more strongly than
positive labels, ASL is particularly suited to highly imbalanced multi-label
classification.

For oversampling, training patents are sampled according to the frequencies of
their assigned CPC subclasses. For a patent with label set $L_i$, the
frequency-based sampling weight is
\begin{equation}
  w_i
  =
  \frac{1}{|L_i|}
  \sum_{j \in L_i}
  \frac{f_{\max}}{f_j},
\end{equation}
where $f_j$ is the training-set frequency of subclass $j$ and
$f_{\max}=\max_j f_j$. Sampling is performed with replacement using
\texttt{WeightedRandomSampler}. We evaluate oversampling strengths of 0.5 and
1.0, which control the influence of the frequency-based weights.

Models are trained for up to four epochs, with gradient accumulation over two
steps and linear warmup over 6\% of the training steps. Within each hyperparameter run, checkpoint selection is based on validation loss. The final hyperparameter configuration is selected according to validation Macro-F1. Table~\ref{tab:longtail_configs} reports the valid
configurations retained from the sweep.

\begin{table}[ht]
\centering
\small
\setlength{\tabcolsep}{6pt}
\renewcommand{\arraystretch}{1.12}
\caption{Validation performance of the BERT-for-Patents configuration sweep.
All configurations use a linear multi-label classification head. Maximum input
length is selected on the validation set within the corresponding training
setting.\label{tab:longtail_configs}
}

\begin{tabular}{llcccc}
\toprule
Configuration
& Loss
& Max length
& Oversampling
& Micro-F1
& Macro-F1 \\
\midrule

BCE baseline
& BCE
& 512
& --
& 0.6180
& 0.4781 \\

Focal
& Focal
& 256
& --
& 0.6258
& 0.5263 \\

Focal + OS
& Focal
& 256
& 0.5
& 0.6215
& 0.5316 \\

\textbf{ASL + OS}
& \textbf{ASL}
& \textbf{256}
& \textbf{0.5}
& \textbf{0.6282}
& \textbf{0.5426} \\

ASL + OS
& ASL
& 256
& 1.0
& 0.6023
& 0.5270 \\

\bottomrule
\end{tabular}

\end{table}

Focal loss improves both Micro-F1 and Macro-F1 relative to the BCE baseline.
Adding moderate oversampling further improves Macro-F1, although with a small
reduction in Micro-F1. The strongest validation configuration combines ASL with
an oversampling strength of 0.5, reaching a Micro-F1 of 0.6282 and a Macro-F1 of
0.5426. Increasing the oversampling strength to 1.0 reduces both metrics. The
ASL configuration with oversampling strength 0.5 is therefore selected as the
long-tail-aware BERT-for-Patents model for subsequent evaluation.

\section{Prompting strategies and templates}
\label{app:prompts}

We evaluate large language models under multiple prompting regimes of increasing
informational content and constraint, reflecting common practical deployment
scenarios. Specifically, we consider: (i) zero-shot prompting; (ii) few-shot
prompting with in-context examples; (iii) constrained prompting based on an
allowed CPC label set, implemented through retrieval from a dictionary of CPC
definitions; and (iv) combinations of few-shot prompting with label-space
constraints. Across all settings, prompts enforce a strict JSON output format to
support deterministic parsing and consistent evaluation.

\subsection{Prompt selection and calibration}

Prior to the main experimental evaluation, we conducted a limited prompt-calibration
phase on the validation set. The purpose of this analysis was to reduce sensitivity to
arbitrary prompt-design choices and to identify a stable configuration for the main
off-the-shelf LLM experiments.

Prompt variants differed in:
(i) instruction wording, including the emphasis placed on recall versus precision; and
(ii) the selection strategy used for few-shot examples.

We evaluated a small set of representative configurations, including neutral and
recall-oriented zero-shot prompts, as well as static, dynamically retrieved, and hybrid
few-shot examples. The hybrid strategy combines a small set of fixed examples with
examples retrieved dynamically from the training set according to their similarity to
the target patent. Prompt selection was performed exclusively on held-out validation
data. No prompt was tuned on the test set, and the selected configuration was
subsequently used unchanged in the main experiments.

Table~\ref{tab:prompt_ablation} reports a representative subset of the validation
results for Qwen3.5-9B. Recall-oriented zero-shot prompting improves substantially
over the neutral zero-shot configuration, particularly in Macro-F1. All few-shot
strategies provide considerably stronger performance than the zero-shot variants.
Dynamic retrieval achieves the highest Micro-F1, whereas the hybrid strategy achieves
the highest Macro-F1 and ties dynamic retrieval for the highest hierarchy-aware F1.
Because the hybrid configuration provides the strongest overall balance across frequent
and infrequent subclasses while preserving the best hierarchy-aware performance, it
was selected as the default few-shot configuration for the main off-the-shelf
Qwen3.5-9B experiments.

\begin{table}[t]
\centering
\caption{
Representative prompt-sensitivity analysis on a held-out validation sample using
Qwen3.5-9B. Recall-oriented zero-shot prompting improves over the neutral
configuration, while all few-shot variants yield substantially stronger performance.
Dynamic retrieval achieves the highest Micro-F1, whereas hybrid retrieval achieves
the highest Macro-F1 and ties dynamic retrieval for the highest hierarchy-aware F1.
The hybrid strategy was therefore selected for the main off-the-shelf Qwen3.5-9B
experiments because it provides the strongest overall balance across the reported
metrics.
}
\label{tab:prompt_ablation}
\small
\begin{tabular}{lccc}
\toprule
Prompt variant & Micro-F1 & Macro-F1 & H-F1 \\
\midrule
Zero-shot (neutral) & 0.285 & 0.037 & 0.345 \\
Zero-shot (high recall) & 0.310 & 0.164 & 0.372 \\
Few-shot (static) & 0.518 & 0.158 & 0.626 \\
Few-shot (dynamic) & 0.540 & 0.246 & 0.637 \\
Few-shot (hybrid) & 0.535 & 0.310 & 0.637 \\
\bottomrule
\end{tabular}

\end{table}

Overall, the calibration analysis shows that both instruction wording and few-shot
example selection materially affect classification performance. Recall-oriented
instructions strengthen the zero-shot baseline, while combining fixed and dynamically
retrieved examples produces the most balanced performance across aggregate,
label-level, and hierarchy-aware metrics. These findings support the use of the hybrid
few-shot strategy as the default prompting configuration for Qwen3.5-9B.

\subsection{Zero-shot prompting}

In the zero-shot setting, the model receives only the patent text and general instructions,
without examples or label constraints.
This represents the cleanest evaluation of the model’s prior knowledge.

\begin{tcolorbox}[breakable, title=Zero-shot system prompt]
\begin{lstlisting}
You are an expert patent examiner for the Cooperative Patent Classification (CPC) system.
Your goal is HIGH RECALL classification: it is worse to miss a relevant CPC subclass
than to include a marginally relevant one.
You assign MULTIPLE CPC subclasses (multi-label) to each patent.
Return ONLY a strict JSON object with a single key "labels",
whose value is a list of CPC subclass codes (e.g. ["G06Q", "Y02D"]).
Return JSON ONLY (no extra text).
\end{lstlisting}
\end{tcolorbox}

\begin{tcolorbox}[breakable, title=Zero-shot user prompt]
\begin{lstlisting}
Classify the following patent into CPC subclasses (4-character level like A01B, G06F, Y02D).

PATENT TEXT:
----------------------------------------
{patent_text}
----------------------------------------

TASK:
- Assign ALL relevant CPC subclasses (multi-label).
- Output between 1 and 7 CPC subclass codes.
- Do NOT invent codes that are not real CPC subclasses.

OUTPUT FORMAT (STRICT):
{"labels": ["G06F", "H04L"]}
\end{lstlisting}
\end{tcolorbox}

\subsection{Few-shot prompting}

In the few-shot setting, the model is additionally provided with labeled patent examples
drawn from the training set. These examples are used to stabilize output format and
encourage multi-label behavior.

\begin{tcolorbox}[breakable, title=Few-shot system prompt]
\begin{lstlisting}
You are an expert patent examiner for the Cooperative Patent Classification (CPC) system.
Your goal is HIGH RECALL classification: it is worse to miss a relevant CPC subclass
than to include a marginally relevant one.
You assign MULTIPLE CPC subclasses (multi-label) to each patent.
Return ONLY a strict JSON object with a single key "labels".
Return JSON ONLY (no extra text).
\end{lstlisting}
\end{tcolorbox}

\begin{tcolorbox}[breakable, title=Few-shot user prompt]
\begin{lstlisting}
You will see some labeled examples first.

FEW-SHOT EXAMPLES:
{fewshot_block}

NOW CLASSIFY THIS PATENT:
----------------------------------------
{patent_text}
----------------------------------------

TASK:
- Assign ALL relevant CPC subclasses (4-character level).
- Output between 1 and 7 CPC subclass codes.

OUTPUT FORMAT (STRICT):
{"labels": ["G06F", "H04L"]}
\end{lstlisting}
\end{tcolorbox}

\subsection{Zero-shot prompting with retrieval-augmented label constraints}

In this setting, we augment zero-shot prompting with external domain knowledge in the
form of CPC subclass definitions.
Specifically, we retrieve a dictionary of relevant CPC subclasses and provide the model
with an explicit \emph{allowed label set}, thereby constraining the output space.
This configuration corresponds to a lightweight retrieval-augmented generation (RAG)
setup, in which no textual evidence is retrieved for the patent itself, but the model is
guided by authoritative classification metadata.

The objective of this variant is twofold: (i) to reduce hallucinated or invalid CPC codes,
and (ii) to encourage recall by making plausible subclasses explicitly available to the
model.
No in-context labeled examples are provided in this setting.

\begin{tcolorbox}[breakable, title=Zero-shot + RAG system prompt]
\begin{lstlisting}
You are an expert patent examiner for the Cooperative Patent Classification (CPC) system.
Your goal is HIGH RECALL classification.
You assign MULTIPLE CPC subclasses to each patent.
You MUST only choose labels from the provided allowed set.
If a CPC subclass is plausibly relevant, you SHOULD include it; do NOT be overly conservative.
Reason hierarchically (Section -> Class -> Subclass), but output ONLY CPC subclass codes.
Return ONLY a strict JSON object with key "labels".
\end{lstlisting}
\end{tcolorbox}

\begin{tcolorbox}[breakable, title=Zero-shot + RAG user prompt]
\begin{lstlisting}
PATENT TEXT:
----------------------------------------
{patent_text}
----------------------------------------

ALLOWED CPC SUBCLASSES (grouped hierarchically):
{allowed_labels}

TASK:
- Assign ALL relevant CPC subclasses.
- Use ONLY codes from the allowed list.
- Return at least one and at most 7 subclasses.

OUTPUT FORMAT (STRICT):
{"labels": ["G06F", "G06Q", "Y02D"]}
\end{lstlisting}
\end{tcolorbox}

\subsection{Few-shot prompting with retrieval-augmented label constraints}

This setting combines in-context learning with retrieval-augmented label constraints.
In addition to the allowed CPC subclass set retrieved from the CPC definition dictionary,
the model is provided with a small number of labeled patent examples drawn from the
training data.

This configuration represents the most informative prompting regime considered in this
work, integrating (i) domain knowledge via CPC definitions and (ii) task-specific
examples illustrating multi-label classification behavior.
As in all other settings, prompts are designed to favor recall and enforce structured
JSON output.

\begin{tcolorbox}[breakable, title=Few-shot + RAG system prompt]
\begin{lstlisting}
You are an expert patent examiner for the Cooperative Patent Classification (CPC) system.
Your goal is HIGH RECALL classification.
You assign MULTIPLE CPC subclasses to each patent.
You MUST only choose labels from the provided allowed set.
If a CPC subclass is plausibly relevant, you SHOULD include it; do NOT be overly conservative.
Reason hierarchically (Section -> Class -> Subclass), but output ONLY CPC subclass codes.
Return ONLY a strict JSON object with key "labels".
\end{lstlisting}
\end{tcolorbox}

\begin{tcolorbox}[breakable, title=Few-shot + RAG user prompt]
\begin{lstlisting}
BELOW ARE SOME EXAMPLES OF HOW TO ASSIGN MULTIPLE CPC SUBCLASSES:
{fewshot_block}
END OF EXAMPLES.

NOW CLASSIFY THE FOLLOWING PATENT:

PATENT TEXT:
----------------------------------------
{patent_text}
----------------------------------------

ALLOWED CPC SUBCLASSES (grouped hierarchically):
{allowed_labels}

INSTRUCTIONS:
- Assign ALL CPC subclasses that are reasonably relevant.
- Prefer OVER-INCLUSION to under-inclusion.
- Use ONLY codes from the allowed list.
- Return at least one and at most 7 subclasses.

OUTPUT FORMAT (STRICT JSON ONLY):
{"labels": ["G06F", "G06Q", "Y02D"]}
\end{lstlisting}
\end{tcolorbox}

\section{Retrieval and RAG Implementation Details}
\label{app:rag_details}

This appendix documents the implementation details of our retrieval-augmented prompting settings used for CPC multi-label prediction.

\subsection{Retrieval corpus: CPC subclass definitions}
\label{app:rag_corpus}
We construct a retrieval corpus from a dictionary of CPC subclass codes mapped to short textual definitions (denoted \texttt{cpc\_labels}). Each retrieval item corresponds to one CPC subclass code (4-character level, e.g., \texttt{G06F}) and its definition. Before embedding, we normalize codes to the 4-character subclass level by (i) uppercasing, (ii) removing any subgroup suffix after ``/'' if present, and (iii) truncating to four characters. Each retrieval item is embedded as a single text string of the form:
\begin{quote}
\texttt{CODE: definition}
\end{quote}
This design ensures that retrieval is performed over the semantic content of the CPC definition while preserving explicit label identity.

\subsection{Bi-encoder retrieval model and embeddings}
\label{app:rag_encoder}
We use a local E5 bi-encoder (\texttt{e5-base-v2}) to embed both queries and CPC-definition passages. Following the E5 convention, we apply explicit prefixes:
\begin{itemize}
    \item \textbf{Query embedding:} patent text prefixed with \texttt{``query: ''}.
    \item \textbf{Passage embedding:} CPC definition text prefixed with \texttt{``passage: ''}.
\end{itemize}
The query text is the concatenation of the patent title and abstract (``\texttt{title. abstract}''), with whitespace normalization.
Tokenization uses truncation to a maximum length of 512 tokens. Embeddings are computed via mean pooling over the last hidden states using the attention mask, and then $\ell_2$-normalized. Because embeddings are normalized, cosine similarity is equivalent to the dot product.

We embed the CPC-definition corpus once (batch size 64) and reuse these vectors for all patents. For patent queries, we embed in batches (default batch size up to 16, capped by the number of items in a batch). All embedding computations run locally (offline), using GPU.

\subsection{Top-$K$ retrieval and allowed-label construction}
\label{app:rag_topk}
For each patent, we compute cosine similarity between the query embedding and all CPC-definition embeddings and retrieve the top-$K$ CPC subclasses, with $K=20$. The resulting list forms the \emph{allowed label set} for that patent. This allowed set is injected into the LLM prompt and also used for strict post-processing: predicted labels not contained in the allowed set are discarded. We fix the size of the retrieved allowed CPC set to $K=20$ across all models and experiments.
We considered alternative sizes of the retrieved allowed label set to assess feasibility.
With $K=10$, the retrieval stage frequently omitted relevant CPC subclasses, leading to
under-inclusive predictions. Conversely, $K=50$ resulted in substantially longer prompts
that approached context-length limits and increased inference cost without qualitative
benefits. Based on these observations, we fixed $K=20$ a priori and did not further tune it.

\subsection{Prompt formatting for RAG settings}
\label{app:rag_prompt}
In RAG prompting, the allowed CPC subclasses are shown together with short definitions and grouped hierarchically to improve readability. Specifically, we group allowed subclasses by CPC \emph{section} (first character) and \emph{class} (first three characters), and list subclasses under each class. Each line is formatted as:
\begin{quote}
\texttt{- CODE --- definition}
\end{quote}
To control prompt length, definitions are truncated to at most 220 characters.

\paragraph{Zero-shot + RAG.}
In the zero-shot + RAG condition, the prompt contains only (i) the patent text and (ii) the retrieved allowed set with definitions. The model is instructed to select all relevant subclasses \emph{from the allowed set} and to return a strict JSON object:
\begin{quote}
\texttt{\{"labels": ["G06F", "H04L"]\}}
\end{quote}
We additionally instruct the model to prefer over-inclusion (high recall) and to output between 1 and 7 labels.

\paragraph{Few-shot + RAG.}
In the few-shot + RAG condition, we use the same allowed-set retrieval as above and prepend few-shot examples to the prompt. Few-shot examples come from the training set and are formatted as \texttt{PATENT TEXT} followed by the corresponding gold label JSON. We include two types of few-shot examples:
\begin{itemize}
    \item \textbf{Static examples:} fixed training row indices (two examples) used in every prompt.
    \item \textbf{Dynamic examples:} retrieved per test instance using TF--IDF cosine similarity over training texts (two examples), excluding the static IDs.
\end{itemize}

We experimented with alternative few-shot configurations (static-only and
dynamic-only retrieval), but observed reduced recall and higher variance across
runs; for this reason, these variants were not pursued further.

\subsection{Few-shot retrieval for examples (TF--IDF)}
\label{app:rag_fewshot_retrieval}
For dynamic few-shot retrieval, we build a TF--IDF representation of training texts (title + abstract) using: maximum \np{50000} features, $(1,2)$-gram range, and minimum document frequency of 2. For each test patent, we compute cosine similarity between its TF--IDF vector and all training vectors and select the top-2 most similar examples (excluding the static examples). Static examples are selected by fixed training indices \texttt{[2, 27]} to stabilize the prompt format across runs.

\subsection{LLM inference settings}
\label{app:rag_llm_inference}
All LLMs are run locally (offline) using Hugging Face Transformers with \texttt{device\_map="auto"}. We use bfloat16 on supported GPUs (otherwise float16). We generate deterministically with greedy decoding (\texttt{do\_sample=False}) and \texttt{max\_new\_tokens=256}. Tokenization uses left padding and truncation to a maximum input length of 2048 tokens.

\subsection{Output parsing, label normalization, and constraint enforcement}
\label{app:rag_parsing}
We enforce structured output by requiring strict JSON in the prompt. In post-processing, we parse the model output as follows:
\begin{enumerate}
    \item Attempt to parse the entire generation as JSON and read the \texttt{"labels"} list.
    \item If this fails, extract the first JSON-like substring using a regex and attempt JSON parsing again.
    \item Normalize each predicted code to the 4-character subclass level (uppercasing, removing subgroup suffix after ``/'', truncating to four characters).
\end{enumerate}
For RAG conditions, we enforce the allowed-label constraint by discarding any normalized label not present in the retrieved allowed set for that patent.

\paragraph{Non-empty fallback.}
Because some LLMs may output invalid JSON or produce empty outputs, we optionally enforce non-empty predictions. If parsing yields no valid labels, we (i) search for CPC-like tokens in the raw text and keep the first token that is in the allowed set; otherwise (ii) fall back to predicting the top-1 retrieved allowed label. In all settings, we cap the final prediction to at most 7 labels per patent. We constrain the number of predicted CPC subclasses to be between 1 and 7.
This range reflects the typical sparsity of CPC subclass annotations in our dataset
and prevents degenerate outputs with excessively many labels, while still allowing
over-inclusion in line with our high-recall objective.

\subsection{Evaluation outputs}
\label{app:rag_outputs}
For each model and setting, we compute micro and macro precision/recall/F$_1$ at the subclass level, together with Acc@1. We also compute hierarchical micro metrics at the section, class, and subclass levels by mapping predicted and gold labels to their corresponding hierarchical prefixes and computing micro precision/recall/F$_1$. Additionally, we export per-label (section/class/subclass) tables including support, predicted positives, TP/FP/FN/TN, precision, recall, F$_1$, and per-label accuracy.
For reproducibility and statistical analysis (e.g., bootstrap), we always save a \texttt{predictions.jsonl} file containing the patent text, gold labels, predicted labels, the full prompt, and retrieval metadata (including the retrieved allowed set and, in the few-shot setting, the IDs of retrieved training examples).

\section{LoRA Fine-Tuning Details}
\label{app:finetuning}

\rev{We fine-tune Qwen3.5-9B-Instruct using parameter-efficient Low-Rank Adaptation (LoRA) for CPC multi-label patent classification. The task is formulated as supervised instruction tuning: given the patent title and abstract, the model is trained to generate a JSON object containing the relevant CPC subclass labels. The training objective is causal language modeling, implemented as token-level cross-entropy loss over the assistant response.}

\rev{All base-model parameters are kept frozen, and only the LoRA adapter weights are updated. Rather than relying on a single manually selected configuration, we conduct a targeted hyperparameter sweep over the LoRA rank, scaling factor, learning rate, and maximum sequence length. The tested configurations include LoRA ranks $r \in \{16,32\}$, scaling factors $\alpha \in \{32,64\}$, learning rates in $\{1 \times 10^{-4}, 2 \times 10^{-4}\}$, and maximum sequence lengths in $\{512,1024,2048\}$ tokens.}

Training examples are formatted using the Qwen chat template. The system prompt instructs the model to act as a patent examiner and to return only a valid JSON object of the form \texttt{\{"labels": [...]\}}. The user message contains the concatenated patent title and abstract, while the assistant message contains the gold CPC subclass labels serialized as JSON. No sequence packing is used.

\rev{The sweep results are reported in Table~\ref{tab:qwen35_lora_sweep}. The strongest configuration in terms of Micro-F1 uses LoRA rank $r=16$, scaling factor $\alpha=32$, a learning rate of $1 \times 10^{-4}$, and a maximum sequence length of 1024 tokens. This configuration reaches a Micro-F1 of 0.6173 and a Macro-F1 of 0.3179. The corresponding best checkpoint is selected at global step 3750, after approximately 1.2 training epochs, on the basis of validation loss.}

\begin{table}[t]
    \centering
    \caption{\rev{Results of the LoRA hyperparameter sweep for Qwen3.5-9B-Instruct. Only Micro-F1 and Macro-F1 are reported.}}
    \label{tab:qwen35_lora_sweep}
    \begin{tabular}{ccccc}
        \toprule
        \rev{LoRA rank} & \rev{LoRA $\alpha$} & \rev{Learning rate} &
        \rev{Maximum length} & \rev{Micro-F1 / Macro-F1} \\
        \midrule
        16 & 32 & $1 \times 10^{-4}$ & 512  & 0.5442 / 0.3012 \\
        16 & 32 & $1 \times 10^{-4}$ & 1024 & \textbf{0.6173} / 0.3179 \\
        16 & 32 & $2 \times 10^{-4}$ & 2048 & 0.5469 / 0.3010 \\
        32 & 64 & $2 \times 10^{-4}$ & 2048 & 0.5523 / 0.3070 \\
        \bottomrule
    \end{tabular}
\end{table}

\rev{Optimization is performed with AdamW. For the selected configuration, the learning rate is $1 \times 10^{-4}$ and the maximum sequence length is 1024 tokens. Within each hyperparameter run, the checkpoint with the lowest validation loss is retained. The final hyperparameter configuration is selected according to
validation Micro-F1. The selected checkpoint achieves an evaluation loss of 0.2597. The final LoRA adapter and tokenizer are saved separately for inference. Training is conducted on a single GPU without model quantization and requires approximately 13.1 hours for the selected run.}

\section{Hybrid routing --- implementation details}
\label{app:hybrid}

The hybrid pipeline combines the long-tail-aware BERT-for-Patents encoder with
the LoRA-adapted Qwen3.5-9B model at inference time. The encoder first predicts
CPC subclasses for all patents and returns calibrated probabilities over the
complete subclass label space. Patents are then ranked according to encoder
uncertainty, and only the most uncertain instances are routed to the LLM. All
non-routed patents retain the original encoder prediction unchanged.

Uncertainty is measured using \emph{max-probability uncertainty}, defined for
patent $i$ as
\[
u_i = 1-\max_j p_{ij},
\]
where $p_{ij}$ is the calibrated encoder probability assigned to subclass $j$.
A high value therefore indicates that the encoder assigns no subclass a
particularly high probability. We evaluate four pre-specified routing fractions:
2\%, 5\%, 10\%, and 20\% of the 10,000-patent test set, corresponding to 200,
500, 1,000, and 2,000 routed patents, respectively. Reporting all four
fractions makes it possible to assess whether encoder--LLM complementarity is
restricted to the most uncertain cases or persists as routing is expanded.

Before fixing the final hybrid prompt, we conducted a limited prompt-calibration
phase on the validation set. The purpose was to ensure that the observed routing
behaviour did not depend on an arbitrary formulation of the correction
instructions. We compared a small number of prompt variants differing in the
emphasis placed on precision versus recall, the treatment of existing encoder
labels, and the instructions governing the addition or removal of candidate
subclasses. Prompt calibration was performed exclusively on validation data,
after which the selected system and user prompts were kept fixed across all
routing fractions and test-set evaluations. Because this calibration was
intended only to stabilise the correction protocol rather than to constitute a
separate experimental comparison, detailed prompt-level results are not
reported.

For routed patents, the LLM operates as a constrained correction mechanism
rather than as an unconstrained classifier. Each prompt contains: (i) the
patent title and abstract; (ii) the encoder's current prediction; (iii) the
25 highest-probability encoder candidates and their probabilities; and
(iv) the available CPC descriptions for the candidate subclasses. The allowed
output set is formed from the encoder's current prediction and its top-ranked
candidate subclasses. Qwen3.5-9B-LoRA is instructed to correct the encoder
prediction using only labels from this set and to return between one and seven
CPC subclasses.

The final pipeline uses replacement-based routing. For each routed patent, a
valid LLM output replaces the encoder prediction in full; predictions for
non-routed patents remain unchanged. Outputs are required to follow a strict
JSON format and are normalised to the four-character CPC subclass level.
Labels outside the allowed candidate set are discarded. If the LLM returns an
empty, invalid, or unparseable output, the pipeline automatically reverts to
the original encoder prediction. This fallback ensures that parsing failures do
not remove the baseline classification.

Generation is deterministic, using greedy decoding with sampling disabled and
a maximum of 64 generated tokens. Predictions and diagnostics are saved
separately for every routing fraction, including the routed patent identifiers,
encoder and hybrid outputs, parsing status, full-test-set metrics, routed-subset
metrics, and rare-label performance.

The final prompt templates, selected through the validation-only calibration
procedure described above, are reported below.

\begin{tcolorbox}[breakable, title=Hybrid routing system prompt]
\begin{lstlisting}
You are an expert patent examiner specialized in Cooperative Patent
Classification (CPC) subclass assignment. You are correcting the
output of an encoder classifier.

Use the patent text, the encoder prediction, the encoder candidate
probabilities, and the available CPC definitions.

Return ONLY valid JSON with exactly one key: "labels".
The value must be a list of CPC subclass codes. Do not explain.

Prefer precision over recall.
Keep encoder labels unless they are clearly unsupported.
Add a new label only when it is strongly supported by the patent
text and the CPC definition.
\end{lstlisting}
\end{tcolorbox}

\begin{tcolorbox}[breakable, title=Hybrid routing user prompt]
\begin{lstlisting}
PATENT TEXT:
----------------------------------------
{patent_text}
----------------------------------------

ENCODER CURRENT PREDICTION:
{encoder_prediction_block}

ENCODER TOP CANDIDATES:
{encoder_candidate_block}

CPC DEFINITIONS:
{cpc_definition_block}

ALLOWED OUTPUT LABELS:
{allowed_labels}

TASK:
Correct the encoder prediction.
- Output the final CPC subclass labels for this patent.
- Use only labels from ALLOWED OUTPUT LABELS.
- Keep encoder labels unless clearly wrong.
- Add labels only when strongly supported by the patent text and
  the CPC definition.
- Return between 1 and 7 labels.
- Return JSON only.

OUTPUT FORMAT:
{"labels": ["G06F", "H04L"]}
\end{lstlisting}
\end{tcolorbox}

\section{Semantic breadth and classification difficulty}
\label{app:semantic_breadth}

Semantic breadth is measured at the CPC-subclass level from the dispersion of patent-text embeddings within each subclass. For each subclass, we compute the distance between every patent representation and the leave-one-out subclass centroid, and use the average distance as a measure of within-class semantic heterogeneity. Higher values therefore indicate that the patents assigned to a subclass occupy a broader and less cohesive semantic space.

For each model, subclass-level F1 is related to semantic breadth, training-set frequency, and test-set support using Spearman rank correlations. Statistical significance is adjusted for multiple comparisons using the false discovery rate (FDR).

We additionally estimate subclass-level regressions of the form
\[
F1_c = \alpha
+ \beta_1 \text{\emph{Breadth}}_c
+ \beta_2 \log(\text{\emph{TrainFreq}}_c)
+ \beta_3 \log(\text{\emph{TestFreq}}_c)
+ \sum_s \gamma_s \text{\emph{Section}}_{cs}
+ \varepsilon_c,
\]
where $c$ indexes CPC subclasses and the section indicators control for broad
differences across CPC domains.

The regressions are estimated separately for the encoder, Qwen3.5-LoRA, and
the selected 5\% hybrid-routing configuration.

Across all configurations, semantic breadth is negatively associated with F1,
whereas training and test frequency are positively correlated with performance.
In the multivariate regressions, the negative breadth coefficient remains
statistically significant after controlling for training frequency, test
frequency, and CPC section. Training frequency also remains positively
associated with F1, whereas test frequency does not retain an independent
association once the other covariates are included. Full correlation and
regression results are reported in Tables~\ref{tab:semantic_correlations} and
\ref{tab:semantic_regressions}.

\begin{table}[t]
\centering
\setlength{\tabcolsep}{4pt}
\resizebox{\linewidth}{!}{%
\begin{tabular}{lcccccc}
\toprule
Model &
$\rho_{\text{breadth}}$ &
$p_{\mathrm{FDR}}^{\text{breadth}}$ &
$\rho_{\text{train freq.}}$ &
$p_{\mathrm{FDR}}^{\text{train}}$ &
$\rho_{\text{test support}}$ &
$p_{\mathrm{FDR}}^{\text{test}}$ \\
\midrule
Encoder
& -0.257 & $1.04\times10^{-5}$
& 0.239 & $2.80\times10^{-5}$
& 0.236 & $2.84\times10^{-5}$ \\
Qwen3.5-LoRA
& -0.235 & $2.84\times10^{-5}$
& 0.287 & $4.18\times10^{-6}$
& 0.267 & $7.32\times10^{-6}$ \\
Hybrid 2\%
& -0.265 & $7.32\times10^{-6}$
& 0.238 & $2.80\times10^{-5}$
& 0.236 & $2.84\times10^{-5}$ \\
Hybrid 5\%
& -0.280 & $4.25\times10^{-6}$
& 0.247 & $1.94\times10^{-5}$
& 0.246 & $1.94\times10^{-5}$ \\
Hybrid 10\%
& -0.269 & $7.32\times10^{-6}$
& 0.244 & $2.07\times10^{-5}$
& 0.246 & $1.94\times10^{-5}$ \\
Hybrid 20\%
& -0.261 & $8.41\times10^{-6}$
& 0.234 & $2.91\times10^{-5}$
& 0.238 & $2.80\times10^{-5}$ \\
\bottomrule
\end{tabular}%
}
\caption{Spearman correlations between subclass-level F1 and semantic breadth,
training frequency, and test support across CPC subclasses. Reported
$p$-values are adjusted for multiple testing using the false discovery rate.}
\label{tab:semantic_correlations}
\end{table}

\begin{table}[t]
\centering
\setlength{\tabcolsep}{4pt}
\resizebox{\linewidth}{!}{%
\begin{tabular}{lccccccccc}
\toprule
Model &
Breadth &
$p_{\mathrm{FDR}}^{\text{breadth}}$ &
Log train freq. &
$p_{\mathrm{FDR}}^{\text{train}}$ &
Log test supp. &
$p_{\mathrm{FDR}}^{\text{test}}$ &
Section Y &
$p_{\mathrm{FDR}}^{Y}$ &
$R^2$ \\
\midrule
Encoder
& -6.076 & $2.79\times10^{-7}$
& 0.086 & 0.039
& -0.006 & 0.995
& -0.349 & $6.69\times10^{-7}$
& 0.325 \\
Qwen3.5-LoRA
& -5.845 & $6.23\times10^{-6}$
& 0.108 & 0.0039
& -0.024 & 0.742
& -0.383 & $4.16\times10^{-7}$
& 0.326 \\
Hybrid 5\%
& -6.447 & $3.93\times10^{-8}$
& 0.088 & 0.032
& -0.006 & 0.995
& -0.346 & $6.69\times10^{-7}$
& 0.344 \\
\bottomrule
\end{tabular}%
}
\caption{Subclass-level regressions of F1 on semantic breadth, log training
frequency, log test support, and CPC-section indicators. The table reports the
encoder, Qwen3.5-LoRA, and the selected 5\% hybrid configuration. Reported
$p$-values are adjusted for multiple testing using the false discovery rate.}
\label{tab:semantic_regressions}
\end{table}
\section{Additional Results}
\label{app:additional_results}

\subsection{Detailed CPC Label Statistics}
\label{app:cpc_statistics}

This subsection provides additional descriptive statistics on the CPC label distribution within the USPTO-70k dataset used in our experiments.

Table~\ref{tab:top_bottom_subclasses} reports the five most frequent and five least frequent CPC subclasses in the training set. The most frequent subclasses correspond to broad and general-purpose technological areas, such as digital data processing and telecommunications, while the least frequent subclasses are highly specialized and represented by fewer than ten training instances.

The ratio between the most frequent and the rarest subclasses exceeds three orders of magnitude, highlighting the severity of the long-tail phenomenon. Such extreme imbalance poses a significant challenge for supervised learning approaches, which tend to optimize performance on frequent labels, and provides further motivation for exploring alternative paradigms such as large language models.

\begin{table}[h]
\centering
\small
\begin{tabular}{l l r p{6.5cm}}
\hline
Group & CPC subclass & Frequency & CPC description \\
\hline
Top-5 
& G06F & 6272 & Electric digital data processing (general-purpose computing, architectures, and data handling) \\
& H04L & 4341 & Transmission of digital information (data communication, networking protocols) \\
& Y10T & 4290 & Technical subjects covered by former US classification (cross-reference technical topics) \\
& H01L & 4168 & Semiconductor devices; electric solid-state devices \\
& H04N & 2592 & Pictorial communication (e.g., television, image and video transmission) \\
\hline
Bottom-5 
& C06C & 9 & Explosives; matches \\
& A44D & 9 & Haberdashery; personal articles not otherwise provided for \\
& B28C & 9 & Working cement, clay, or stone (mixing, shaping, or processing) \\
& C10C & 9 & Processing of coal or petroleum products (e.g., destructive distillation) \\
& C12L & 4 & Pitching or depitching machines; cellar tools \\
\hline
\end{tabular}
\caption{Top-5 and Bottom-5 CPC subclasses by frequency in the training set, with official CPC subclass descriptions.}
\label{tab:top_bottom_subclasses}
\end{table}

\begin{figure}[h]
    \centering
    \includegraphics[width=0.7\textwidth]{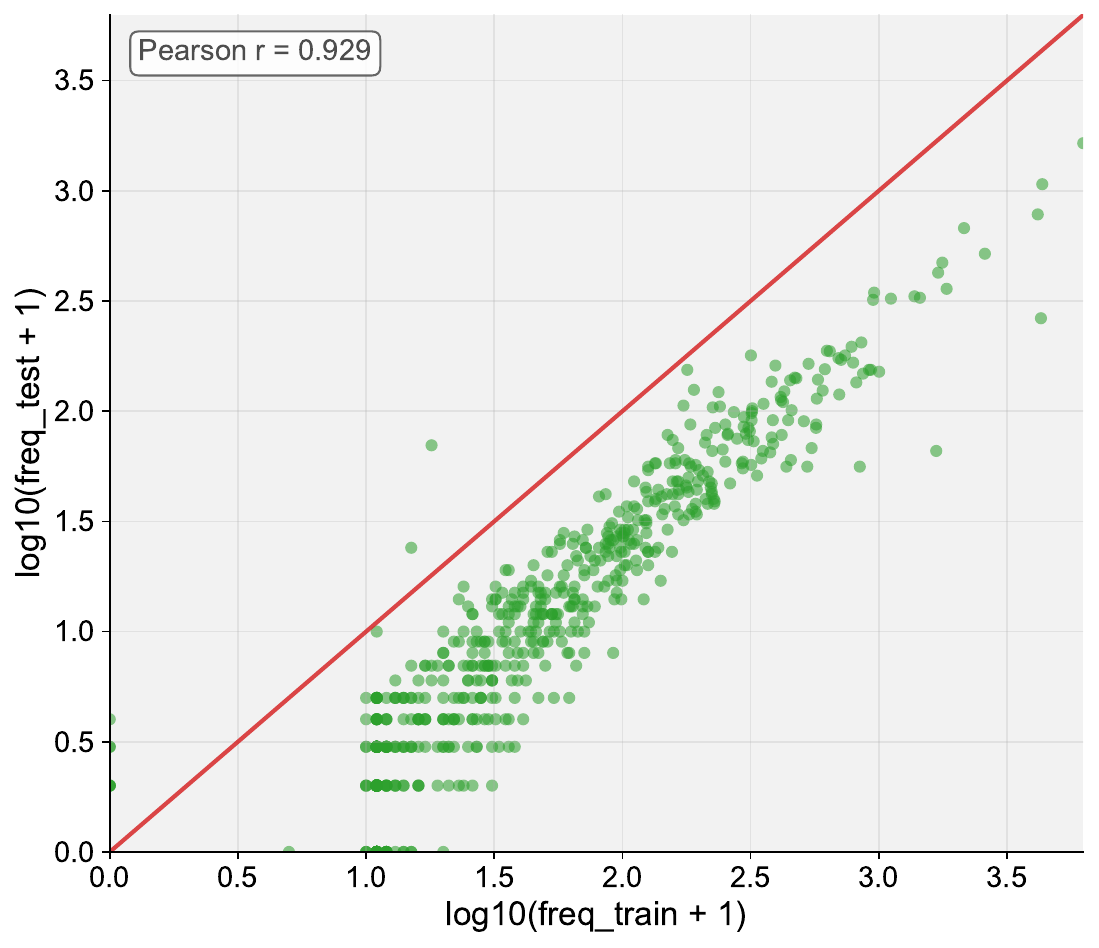}
    \caption{
    Correlation between CPC subclass frequencies in the training and test sets.
    Each point corresponds to a CPC subclass.
    Frequencies are shown on a logarithmic scale ($\log_{10}(\mathrm{count}+1)$) to account for the long-tailed distribution of labels.
    The strong linear relationship indicates that the train–test split preserves the original subclass imbalance, including rare and emerging technologies.
    }
    \label{fig:train_test_freq_corr_appendix}
\end{figure}

Finally, Figure~\ref{fig:train_test_freq_corr_appendix} reports the correlation between subclass frequencies in the training and test sets. The strong positive correlation indicates that the test set largely preserves the label distribution observed during training, suggesting that differences in model performance are not driven by distributional shifts between splits.

\begin{table}[!htbp]
\centering
\small
\setlength{\tabcolsep}{6pt}
\renewcommand{\arraystretch}{1.15}

\scalebox{0.72}{
\begin{tabular}{@{}clcccc@{}}
\toprule
\textbf{Experiment} &
\multicolumn{1}{c}{\textbf{Model / configuration}} &
\textbf{Micro-F1} &
\textbf{Macro-F1} &
\textbf{H-F1} &
\textbf{Bootstrap CI (Micro-F1)} \\
\midrule

\multirow{4}{3.4cm}{\emph{Baseline encoders: full label space}} &
BERT               & 0.401 & 0.044 & 0.494 & [0.394, 0.407] \\
&SciBERT            & 0.439 & 0.059 & 0.527 & [0.432, 0.446] \\
&PatentSBERTa       & 0.328 & 0.016 & 0.419 & [0.322, 0.335] \\
&BERT-for-Patents   & 0.592 & 0.208 & 0.673 & [0.585, 0.597] \\
\midrule

\multirow{4}{3.4cm}{\emph{Baseline encoders: constrained label space}} &
BERT (constrained)              & 0.303 & 0.053 & 0.433 & [0.296, 0.309] \\
&SciBERT (constrained)           & 0.330 & 0.064 & 0.451 & [0.322, 0.337] \\
&PatentSBERTa (constrained)      & 0.260 & 0.033 & 0.393 & [0.253, 0.266] \\
&BERT-for-Patents (constrained)  & 0.449 & 0.095 & 0.551 & [0.442, 0.456] \\
\midrule

\multirow{5}{3.4cm}{\emph{Off-the-shelf LLMs: zero-shot, no RAG}}
& Phi-3 Mini               & 0.112 & 0.046 & 0.225 & [0.108, 0.118] \\
& Mistral-7B-Instruct      & 0.203 & 0.044 & 0.306 & [0.198, 0.209] \\
& LLaMA-3.1-8B-Instruct    & 0.245 & 0.107 & 0.347 & [0.240, 0.251] \\
& Qwen3.5-9B               & 0.455 & 0.164 & 0.547 & [0.450, 0.462] \\
& Gemma4-12B               & 0.389 & 0.049 & 0.543 & [0.383, 0.394] \\
\midrule

\multirow{5}{3.4cm}{\emph{Off-the-shelf LLMs: few-shot, no RAG}}
&Phi-3 Mini               & 0.102 & 0.026 & 0.192 & [0.097, 0.106] \\
&Mistral-7B-Instruct      & 0.126 & 0.034 & 0.215 & [0.123, 0.133] \\
&LLaMA-3.1-8B-Instruct    & 0.206 & 0.103 & 0.294 & [0.201, 0.213] \\
&Qwen3.5-9B               & 0.525 & 0.320 & 0.635 & [0.520, 0.530] \\
&Gemma4-12B               & 0.464 & 0.252 & 0.601 & [0.459, 0.468] \\
\midrule

\multirow{5}{3.4cm}{\emph{Off-the-shelf LLMs: zero-shot + RAG}}
& Phi-3 Mini               & 0.130 & 0.101 & 0.270 & [0.126, 0.134] \\
& Mistral-7B-Instruct      & 0.175 & 0.135 & 0.323 & [0.172, 0.181] \\
& LLaMA-3.1-8B-Instruct    & 0.189 & 0.144 & 0.344 & [0.186, 0.196] \\
& Qwen3.5-9B               & 0.358 & 0.227 & 0.499 & [0.355, 0.362] \\
& Gemma4-12B               & 0.356 & 0.228 & 0.502 & [0.351, 0.359] \\
\midrule

\multirow{5}{3.4cm}{\emph{Off-the-shelf LLMs: few-shot + RAG}}
& Phi-3 Mini               & 0.139 & 0.125 & 0.292 & [0.133, 0.143] \\
& Mistral-7B-Instruct      & 0.169 & 0.112 & 0.317 & [0.159, 0.170] \\
& LLaMA-3.1-8B-Instruct    & 0.177 & 0.131 & 0.328 & [0.170, 0.181] \\
& Qwen3.5-9B               & 0.387 & 0.246 & 0.528 & [0.381, 0.393] \\
& Gemma4-12B               & 0.402 & 0.259 & 0.534 & [0.397, 0.407] \\

\midrule

\multirow{2}{3.4cm}{\emph{Task-adapted models}}
& BERT-for-Patents + long-tail training
  & 0.616 & 0.410 & 0.694 & [0.608, 0.618] \\
& Qwen3.5-9B + LoRA
  & 0.602 & 0.379 & 0.686 & [0.598, 0.606] \\
\midrule

\multirow{4}{3.4cm}{\emph{Hybrid encoder--LLM routing: full test set}}
& BERT-for-Patents + Qwen-LoRA (2\%)
  & 0.615 & 0.411 & 0.694 & [0.611, 0.619] \\
& BERT-for-Patents + Qwen-LoRA (5\%)
  & 0.614 & 0.412 & 0.695 & [0.610, 0.617] \\
& BERT-for-Patents + Qwen-LoRA (10\%)
  & 0.613 & 0.411 & 0.695 & [0.609, 0.617] \\
& BERT-for-Patents + Qwen-LoRA (20\%)
  & 0.608 & 0.410 & 0.693 & [0.604, 0.612] \\
\midrule

\multirow{8}{3.4cm}{\emph{Hybrid routing: routed subsets only}}
& Encoder (2\%)
  & 0.210 & 0.058 & 0.344 & [0.207, 0.214] \\
& Hybrid (2\%)
  & 0.338 & 0.122 & 0.477 & [0.332, 0.343] \\

& Encoder (5\%)
  & 0.298 & 0.121 & 0.424 & [0.294, 0.302] \\
& Hybrid (5\%)
  & 0.378 & 0.183 & 0.516 & [0.374, 0.382] \\

& Encoder (10\%)
  & 0.377 & 0.200 & 0.492 & [0.373, 0.381] \\
& Hybrid (10\%)
  & 0.418 & 0.247 & 0.544 & [0.413, 0.422] \\

& Encoder (20\%)
  & 0.448 & 0.283 & 0.552 & [0.444, 0.452] \\
& Hybrid (20\%)
  & 0.451 & 0.301 & 0.569 & [0.449, 0.455] \\
\bottomrule
\end{tabular}
}

\caption{Complete predictive performance results for CPC subclass multi-label
classification. Baseline encoders are supervised models trained with standard
binary cross-entropy. Constrained encoder variants restrict predictions to a retrieved
candidate label set. Off-the-shelf LLMs are evaluated under zero-shot, few-shot,
retrieval-augmented, and combined prompting regimes. Task-adapted models include
long-tail-aware encoder configurations and LoRA-adapted LLMs. The revised hybrid
system combines long-tail-aware BERT-for-Patents with Qwen3.5-9B-LoRA through
uncertainty-based routing.}
\label{tab:appendix_full_performance}
\end{table}

\newcolumntype{H}{>{\setbox0=\hbox\bgroup}c<{\egroup}@{}}

\begin{table}[t]
\centering
\small
\setlength{\tabcolsep}{4pt}
\renewcommand{\arraystretch}{1.15}

\scalebox{0.58}{
\begin{tabular}{@{}Hllcccccc@{}}
\toprule
\textbf{Stage} & \textbf{Setting} & \textbf{Model} &
\textbf{Time (min)} & \textbf{Energy (kWh)} & \textbf{CO$_2$ (kg)} &
\textbf{Time/pat. (s)} & \textbf{Energy/pat. (Wh)} &
\textbf{CO$_2$/pat. (g)} \\
\midrule

\multicolumn{9}{c}{\emph{TRAINING}}\\
\midrule

\multirow{5}{*}{Training}
& \multirow{4}{*}{Encoder} & BERT
& 12 & 0.074 & $3.0\times10^{-4}$ &
\multicolumn{3}{c}{--------------------- \emph{does not apply} ---------------------} \\

& & PatentSBERTa
& 14 & 0.086 & $3.0\times10^{-4}$ &
\multicolumn{3}{c}{--------------------- \emph{does not apply} ---------------------} \\

& & SciBERT
& 12 & 0.074 & $3.0\times10^{-4}$ &
\multicolumn{3}{c}{--------------------- \emph{does not apply} ---------------------} \\

& & BERT-for-Patents
& 14 & 0.095 & $3.9\times10^{-4}$ &
\multicolumn{3}{c}{--------------------- \emph{does not apply} ---------------------} \\

\cmidrule(lr){3-9}

& LoRA & Qwen3.5-9B
& 786 & 0.648 & $2.5\times10^{-3}$ &
\multicolumn{3}{c}{--------------------- \emph{does not apply} ---------------------} \\

\midrule

\multicolumn{9}{c}{\emph{INFERENCE}}\\
\midrule

\multirow{25}{*}{Inference}

& \multirow{4}{*}{Encoders} & BERT
& 1 & 0.010 & $3.9\times10^{-5}$ &
$8.4\times10^{-3}$ & $1.0\times10^{-3}$ & $3.9\times10^{-6}$ \\

& & PatentSBERTa
& 2 & 0.011 & $4.2\times10^{-5}$ &
$9.1\times10^{-3}$ & $1.1\times10^{-3}$ & $4.2\times10^{-6}$ \\

& & SciBERT
& 1 & 0.010 & $3.8\times10^{-5}$ &
$8.2\times10^{-3}$ & $9.5\times10^{-4}$ & $3.8\times10^{-6}$ \\

& & BERT-for-Patents
& 1 & 0.010 & $4.0\times10^{-5}$ &
$8.2\times10^{-3}$ & $1.0\times10^{-3}$ & $4.0\times10^{-6}$ \\

\cmidrule(lr){3-9}

& \multirow{4}{*}{Few-shot} & LLaMA
& 174 & 1.111 & $4.4\times10^{-3}$ &
1.0 & $1.1\times10^{-1}$ & $4.4\times10^{-4}$ \\

& & Mistral
& 91 & 0.575 & $2.3\times10^{-3}$ &
0.5 & $6.0\times10^{-2}$ & $2.3\times10^{-4}$ \\

& & Phi
& 37 & 0.215 & $8.5\times10^{-4}$ &
0.2 & $2.0\times10^{-2}$ & $8.5\times10^{-5}$ \\

& & Qwen
& 65 & 0.478 & $1.9\times10^{-3}$ &
0.4 & $5.0\times10^{-2}$ & $1.9\times10^{-4}$ \\

& & Gemma
& 118 & 0.689 & $2.9\times10^{-3}$ &
0.7 & $6.89\times10^{-2}$ & $2.9\times10^{-4}$ \\

\cmidrule(lr){3-9}

& \multirow{4}{*}{Zero-shot + RAG} & LLaMA
& 115 & 0.784 & $3.1\times10^{-3}$ &
0.7 & $8.0\times10^{-2}$ & $3.1\times10^{-4}$ \\

& & Mistral
& 113 & 0.807 & $3.2\times10^{-3}$ &
0.7 & $8.0\times10^{-2}$ & $3.2\times10^{-4}$ \\

& & Phi
& 36 & 0.243 & $9.6\times10^{-4}$ &
0.2 & $2.0\times10^{-2}$ & $9.6\times10^{-5}$ \\

& & Qwen
& 93 & 0.572 & $2.3\times10^{-3}$ &
0.6 & $5.7\times10^{-2}$ & $2.3\times10^{-4}$ \\

& & Gemma
& 1196 & 7.440 & $2.9\times10^{-2}$ &
7.1 & $7.4\times10^{-1}$ & $2.9\times10^{-3}$ \\

\cmidrule(lr){3-9}

& & Qwen3.5-9B + LoRA
& 101 & 0.583 & $2.3\times10^{-3}$ &
0.6 & $5.8\times10^{-2}$ & $2.3\times10^{-4}$ \\

\cmidrule(lr){3-9}

& \multirow{5}{*}{Hybrid routing}

& BERT-for-Patents long-tail + Qwen3.5-LoRA (2\%)
& 4 & 0.027 & $1.1\times10^{-4}$ &
$2.5\times10^{-2}$ & $2.7\times10^{-3}$ & $1.1\times10^{-5}$ \\

& & BERT-for-Patents long-tail + Qwen3.5-LoRA (5\%)
& 10 & 0.064 & $2.6\times10^{-4}$ &
$6.0\times10^{-2}$ & $6.4\times10^{-3}$ & $2.6\times10^{-5}$ \\

& & BERT-for-Patents long-tail + Qwen3.5-LoRA (10\%)
& 19 & 0.122 & $4.9\times10^{-4}$ &
0.1 & $1.2\times10^{-2}$ & $4.9\times10^{-5}$ \\

& & BERT-for-Patents long-tail + Qwen3.5-LoRA (20\%)
& 33 & 0.196 & $7.8\times10^{-4}$ &
0.2 & $2.0\times10^{-2}$ & $7.8\times10^{-5}$ \\

\bottomrule
\end{tabular}
}

\caption{Computational cost measured with CodeCarbon for training and inference.
Inference results are totals per evaluation run on \textit{N}=\np{10000} patents.
}
\label{tab:cost_time_energy_co2}
\end{table}

\rev{\subsection{Hybrid routing: full test set vs.\ routed patents}
\label{app:wilcoxon_hybrid_routed}

To assess whether uncertainty-based routing improves performance relative to the encoder baseline, we perform paired Wilcoxon signed-rank tests on per-subclass F1 scores. For each routing threshold, the hybrid model is compared with the encoder on two evaluation samples: the full test set and the subset of patents actually routed to Qwen. Tests are computed over matched CPC subclasses and complemented with rank-biserial correlations ($r$), where positive values indicate higher F1 for the hybrid model. Bonferroni-adjusted $p$-values are reported within each family of routing-threshold comparisons.

Table~\ref{tab:wilcoxon_hybrid_routed} shows that the conclusions differ sharply between the full test set and the routed subset. Across the full test set, none of the hybrid configurations yields a statistically significant subclass-level improvement over the encoder after correction. Mean changes are small, ranging from $-0.001$ to $+0.002$, and the corresponding effect sizes are negligible or small. This reflects the fact that only a limited share of patents is reassigned by the hybrid system, leaving most subclass-level predictions unchanged.

\begin{table*}[t]
\centering
\small
\setlength{\tabcolsep}{7pt}
\renewcommand{\arraystretch}{1.15}

\begin{tabular}{@{}llrrr@{}}
\toprule
\textbf{Routing fraction} &
\textbf{Evaluation sample} &
\textbf{$\Delta$ Micro-F1} &
\textbf{$p_{\mathrm{Bonf.}}$} &
\textbf{$r$} \\
\midrule

2\% &
Full test set &
-0.001 &
0.305 &
0.227 \\

2\% &
Routed patents only &
+0.128 &
$<0.001$ &
0.813 \\

5\% &
Full test set &
-0.002 &
1.000 &
0.062 \\

5\% &
Routed patents only &
+0.080 &
$<0.001$ &
0.572 \\

10\% &
Full test set &
-0.003 &
1.000 &
-0.036 \\

10\% &
Routed patents only &
+0.041 &
0.004 &
0.315 \\

20\% &
Full test set &
-0.008 &
0.133 &
-0.156 \\

20\% &
Routed patents only &
+0.003 &
1.000 &
-0.047 \\

\bottomrule
\end{tabular}

\caption{
Comparison between hybrid routing and the encoder baseline on the full test set
and on the patents selected for routing at each routing fraction.
$\Delta$ Micro-F1 denotes the difference in aggregate Micro-F1 between the hybrid
and encoder predictions. Statistical significance is assessed using paired Wilcoxon
signed-rank tests on per-subclass F1 scores. Positive rank-biserial correlations ($r$)
indicate that subclass-level F1 tends to be higher for the hybrid model.
$p_{\mathrm{Bonf.}}$ denotes Bonferroni-adjusted $p$-values.
}
\label{tab:wilcoxon_hybrid_routed}
\end{table*}

The routed-subset analysis provides direct evidence of selective encoder--LLM complementarity. At routing fractions of 2\%, 5\%, and 10\%, the hybrid significantly outperforms the encoder on the patents selected by the routing mechanism. The mean subclass-level improvement decreases monotonically from $\Delta$F1$=0.075$ at 2\% routing to $0.052$ at 5\% and $0.023$ at 10\%. The corresponding rank-biserial correlations decline from a large effect at 2\% ($r=0.813$) to a large effect at 5\% ($r=0.572$) and a medium effect at 10\% ($r=0.315$).

At 20\% routing, this advantage disappears. The mean change on the routed subset is effectively zero, more subclasses worsen than improve, and the effect size becomes negligible ($r=-0.047$). This pattern indicates that the routing mechanism is most effective when restricted to a small set of highly uncertain patents. As the routing threshold expands, progressively less suitable cases are passed to Qwen, diluting and eventually eliminating the gains observed at lower routing fractions.
}

\subsection{External validation on EPO patents}
\label{app:epo_results}

To assess external validity, we evaluate the final models on \np{10000} EPO
patents published between 2019 and 2022. The models are trained on USPTO data
and applied without retraining.

Table~\ref{tab:epo_performance} confirms the main findings. Adapted
BERT-for-Patents improves substantially over the base encoder, while LoRA
adaptation strengthens Qwen3.5 without allowing it to surpass the encoder. The
5\% hybrid achieves nearly identical Micro-F1 to the adapted encoder, with small
gains in Macro-F1 and hierarchical F1.

On the routed subset, however, Micro-F1 increases from 0.264 to 0.399
($p=1.68\times10^{-21}$; rank-biserial $r=0.803$). This confirms that hybrid
routing is most useful as a targeted refinement mechanism for uncertain patents.

\begin{table*}[t]
\centering
\small
\setlength{\tabcolsep}{5pt}
\caption{Performance on \np{10000} EPO patents. Rare F1 is computed over
subclasses in the bottom 20\% of the USPTO training-frequency distribution.}
\label{tab:epo_performance}
\begin{tabular}{lccccc}
\toprule
Model & Micro-F1 & Macro-F1 & H-F1 & Rare F1 & Section Y Micro-F1 \\
\midrule
BERT-for-Patents base
& 0.5313 & 0.2018 & 0.6197 & 0.0000 & 0.3783 \\
BERT-for-Patents adapted
& 0.5820 & 0.3898 & 0.6620 & 0.1801 & 0.4187 \\
Qwen3.5 few-shot
& 0.4801 & 0.2695 & 0.5948 & 0.0454 & 0.1622 \\
Qwen3.5-LoRA
& 0.5281 & 0.3395 & 0.6259 & 0.1048 & 0.2780 \\
Hybrid 5\%
& 0.5822 & 0.3973 & 0.6648
& 0.1930 & 0.4164 \\
\bottomrule
\end{tabular}
\end{table*}

\begin{table}[t]
\centering
\small
\caption{Performance on the 5\% routed EPO subset.}
\label{tab:epo_routed_subset}
\begin{tabular}{lc}
\toprule
Measure & Value \\
\midrule
Encoder Micro-F1 & 0.2638 \\
Hybrid Micro-F1 & 0.3987 \\
Wilcoxon $p$-value & $1.68\times10^{-21}$ \\
Rank-biserial correlation & 0.803 \\
\bottomrule
\end{tabular}
\end{table}

\end{appendices}


\end{document}